\documentclass[11pt]{article}
\usepackage{amssymb,amsmath,amsfonts}
\usepackage{graphicx}
\usepackage{graphics}
\usepackage{eepic,epsfig}

\makeatletter
\@addtoreset{equation}{section}

\makeatother

\begin{document}

\title{Cherenkov radiation and emission of surface polaritons \\
from charges moving paraxially outside \\
a dielectric cylindrical waveguide}
\author{A. A. Saharian$^{1,2}$\thanks{%
E-mail: saharian@ysu.am}, L. Sh. Grigoryan$^{1}$, A. Kh. Grigorian$^{1}$,
\and H. F. Khachatryan$^{1}$, A. S. Kotanjyan$^{1,2}$ \vspace{0.3cm} \\
\textit{$^1$Institute of Applied Problems of Physics NAS RA, }\\
\textit{25 Hr. Nersessian Str., 0014 Yerevan, Armenia} \vspace{0.3cm}\\
\textit{$^2$ Department of Physics, Yerevan State University,}\\
\textit{1 Alex Manoogian Street, 0025 Yerevan, Armenia }}
\maketitle

\begin{abstract}
We investigate the radiation from a charged particle moving outside a
dielectric cylinder parallel to its axis. It is assumed that the cylinder is
immersed into a homogeneous medium. The expressions are given for the vector
potential and for the electric and magnetic fields. The spectral
distributions are studied for three types of the radiations: (i) Cherenkov
radiation (CR) in the exterior medium, (ii) radiation on the guided modes of
the dielectric cylinder, and (iii) emission of surface polaritons. Unlike
the first two types of radiations, there is no velocity threshold for the
generation of surface polaritons. The corresponding radiation is present in
the spectral range where the dielectric permittivities of the cylinder and
surrounding medium have opposite signs. The spectral range of the emitted
surface polaritons becomes narrower with decreasing energy of the particle.
The general results are illustrated for a special case of the Drude model
for dispersion of the dielectric permittivity of the cylinder. We show that
the presence of the cylinder may lead to the appearance of strong narrow
peaks in the spectral distribution of the CR in the exterior medium. The
conditions are specified for the appearance of those peaks and the
corresponding heights and widths are analytically estimated. The collective
effects of particles in bunches are discussed.
\end{abstract}

\bigskip

\section{Introduction}

The polarization of a medium by moving charged particles gives rise to a
number of radiation processes. Examples are the Cherenkov radiation (CR),
transition radiation and the diffraction radiation. Among those radiation
processes, the remarkable properties of the CR (for reviews see \cite%
{Jell58,Bolo57}) have resulted in a wide variety of applications, including
counting and identifying of high-energy particles, cosmic-ray physics,
high-power radiation sources in various spectral ranges, particle
accelerating systems, medical imaging and therapy and so on. These
applications motivate the importance of the further investigations for
various mechanisms to control the spectral and angular characteristics of
the radiation intensity. In particular, recent advances in nanophysics,
photonic crystals and metamaterials provide new possibilities for the CR
manipulations. Technologies are available that allow to design materials
with specified electric and magnetic properties, including the dispersion
relations for effective dielectric permittivity and magnetic permeability
\cite{Marq08}. An exciting possibility is that the permittivity and
permeability can be made simultaneously negative in some frequency range
(double-negative or left-handed metamaterials). In that spectral range the
wave vector and the electromagnetic field vectors form a left-handed system
and the CR is emitted in the backward direction with respect to the velocity
of the charged particle (reversed CR) \cite{Vese68} (for reviews see \cite%
{Chen11}). Significant progress in metamaterial-related research has
stimulated active theoretical and experimental investigations of the
reversed CR (see \cite{Lu03,Fern12} and references therein).

From the point of view of the CR characteristics control, another important
area of research is the investigation of the influence of interfaces of
media with different electrodynamical properties. The previous
considerations of the effects include planar, cylindrical and spherical
boundaries (for reviews of early research see \cite{Jell58,Bolo57}). More
complicated geometries and approximate methods for evaluation of the
radiation fields and intensity have been considered in \cite{Garc02}. The CR
from a short relativistic electron bunch in dielectric loaded waveguides
with different periodic structures is a promising candidate for high power
narrow band-width source with adjustable spectral range (see, for instance,
\cite{Wals77,Cook09,Galy14} and references therein). Various amplification
mechanisms have been discussed. The Cherenkov emission of surface waves in
planar structures has been considered in \cite{Baku09}. The authors of \cite%
{Liu12} investigate the CR emitted by surface plasmon polaritons.

In the present paper we consider the CR and the emission of guided modes and
surface polaritons by charged particles moving outside a cylindrical
dielectric waveguide, parallel to its axis (for various aspects of
interactions of charged particles with cylindrical structures see \cite%
{Zaba01} and references given there). Exact analytical expressions are
provided for the spectral distributions of all these types of radiations.
The conditions are specified under which the cylinder can essentially
influence the spectral density of the CR in the surrounding medium. Aside
from applications as a source of the electromagnetic radiation in various
spectral regions, the results presented can be used to test the accuracy of
various approximate methods used for investigation of the CR in more
complicated geometries of interfaces. The properties of the emitted surface
polaritons are highly sensitive to the geometry of the surface and this
offers an alternative surface probe. Among important physical realizations
of cylindrical waveguides, with radii tunable in relatively wide range, are
metallic and semiconductor carbon nanotubes.

The layout of the paper is as follows. In the next section, expressions for
the vector potential and for the electric and magnetic fields are provided.
Assuming that the Cherenkov condition in the exterior medium is satisfied,
in section \ref{sec:Cher} a formula is derived for the spectral density of
the radiation evaluating the energy flux through a cylindrical surface with
large radius. The features of the radiation intensity are described
depending on the relative permittivity. The energy losses are investigated
in section \ref{sec:Losses}. An alternative expression is provided for the
spectral density of the CR in the exterior medium. The radiation on the
guided modes of the dielectric cylinder is discussed in section \ref%
{sec:Guiding}. The radiation intensity for surface polaritons is considered
in section \ref{sec:SP}. Section \ref{sec:Conc} concludes the main results
of the paper.

\section{Electromagnetic fields}

\label{sec:Fields}

\begin{figure}[tbph]
\begin{center}
\epsfig{figure=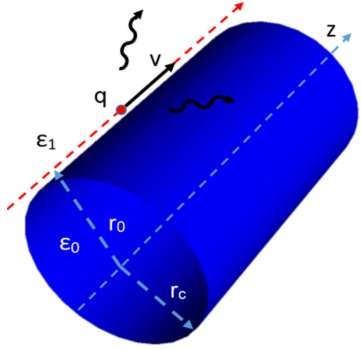,width=5.cm,height=5.2cm}
\end{center}
\caption{The problem geometry and the notations.}
\label{fig1}
\end{figure}

Consider a point charge $q$ moving parallel to the axis of a cylinder with
dielectric permittivity $\varepsilon _{0}$ and with the radius $r_{c}$. The
distance of the charge trajectory from the axis will be denoted by $%
r_{0}>r_{c}$ and it will be assumed that the cylinder is immersed into a
homogeneous medium with dielectric permittivity $\varepsilon _{1}$ (see
figure \ref{fig1}, the magnetic permeabilities for both the cylinder and
surrounding medium will be taken to be unit). In accordance with the problem
symmetry we will use cylindrical coordinates $(r,\phi ,z)$ with the axis $z$
along the axis of the cylinder. In the generalized Lorentz gauge, the vector
potential of the electromagnetic field created by the charge is expressed in
terms of the electromagnetic field Green tensor $G_{il}(\mathbf{r},t,\mathbf{%
r}^{\prime },t^{\prime })$ as
\begin{equation}
A_{i}(t,\mathbf{r})=-\frac{1}{2\pi ^{2}c}\int dt^{\prime }d\mathbf{r}%
^{\prime }\sum_{l=1}^{3}G_{il}(t,\mathbf{r},t^{\prime },\mathbf{r}^{\prime
})j_{l}(t^{\prime },\mathbf{r}^{\prime }),  \label{vecpot}
\end{equation}%
where $j_{l}(t,\mathbf{r})$ is the current density for the source. In the
problem under consideration the only nonzero component of the latter is
given by
\begin{equation}
j_{3}(t,\mathbf{r})=\frac{q}{r}v\delta (r-r_{0})\delta (\phi -\phi
_{0})\delta (z-vt),  \label{jz}
\end{equation}%
with $v$ being the charge velocity.

It is convenient to write the relation (\ref{vecpot}) in terms of the
partial Fourier components $A_{l,n}(k_{z},r)$ of the vector potential
defined in accordance with%
\begin{equation}
A_{l}(t,\mathbf{r})=\sum_{n=-\infty }^{\infty }e^{in(\phi -\phi
_{0})}\int_{-\infty }^{\infty }dk_{z}\,e^{ik_{z}(z-vt)}A_{l,n}(k_{z},r).
\label{vecF}
\end{equation}%
By using the Fourier expansion%
\begin{equation}
G_{il}(t,\mathbf{r},t^{\prime },\mathbf{r}^{\prime })=\sum_{n=-\infty
}^{\infty }\int_{-\infty }^{\infty }d\omega \int_{-\infty }^{\infty
}dk_{z}\,G_{il,n}(\omega ,k_{z},r,r^{\prime })e^{in(\phi -\phi ^{\prime
})+ik_{z}(z-z^{\prime })-i\omega (t-t^{\prime })},  \label{GFour}
\end{equation}%
from (\ref{vecpot}) we get%
\begin{equation}
A_{l,n}(k_{z},r)=-\frac{qv}{\pi c}G_{l3,n}(vk_{z},k_{z},r,r_{0}).
\label{Anl}
\end{equation}

In \cite{Grig95} a recurrence scheme was developed for evaluation of the
Green tensor in a medium with an arbitrary number of cylindrically symmetric
homogeneous layers. In the problem at hand the Green tensor is obtained by
using the corresponding tensor in a homogeneous medium. In particular, for
the region $r>r_{0}$ the Fourier components of the Green tensor appearing in
(\ref{Anl}) are given by the expressions \cite{Grig95}
\begin{eqnarray}
G_{l3,n}(\omega ,k_{z},r,r_{0}) &=&\frac{i^{2-l}k_{z}}{2r_{c}}J_{n}(\lambda
_{0}r_{c})\frac{H_{n}(\lambda _{1}r_{0})}{\alpha _{n}V_{n}^{H}}\sum_{p=\pm
1}p^{l-1}J_{n+p}(\lambda _{0}r_{c})\frac{H_{n+p}(\lambda _{1}r)}{V_{n+p}^{H}}%
,  \notag \\
G_{33,n}(\omega ,k_{z},r,r_{0}) &=&\frac{\pi }{2i}\left[ J_{n}(\lambda
_{1}r_{0})-H_{n}(\lambda _{1}r_{0})\frac{V_{n}^{J}}{V_{n}^{H}}\right]
H_{n}(\lambda _{1}r),  \label{GFz}
\end{eqnarray}%
where $l=1,2$, and $\lambda _{j}^{2}=\omega ^{2}\varepsilon
_{j}/c^{2}-k_{z}^{2}$ with $j=0,1$. In (\ref{GFz}), $J_{n}(x)$ is the Bessel
function, $H_{n}(x)=H_{n}^{(1)}(x)$ is the Hankel function of the first
kind, and we have introduced the notation
\begin{equation}
V_{n}^{F}=J_{n}(\lambda _{0}r_{c})\partial _{r_{c}}F_{n}(\lambda
_{1}r_{c})-[\partial _{r_{c}}J_{n}(\lambda _{0}r_{c})]F_{n}(\lambda
_{1}r_{c}),  \label{VF}
\end{equation}%
for $F=J,H$. The function $\alpha _{n}$ in the expression for the component $%
G_{l3,n}(\omega ,k_{z},r,r^{\prime })$ is given by the formula
\begin{equation}
\alpha _{n}=\frac{\varepsilon _{0}}{\varepsilon _{1}-\varepsilon _{0}}+\frac{%
1}{2}\sum_{l=\pm 1}\left[ 1-\frac{\lambda _{1}}{\lambda _{0}}\frac{%
J_{n+l}(\lambda _{0}r_{c})H_{n}(\lambda _{1}r_{c})}{J_{n}(\lambda
_{0}r_{c})H_{n+l}(\lambda _{1}r_{c})}\right] ^{-1}.  \label{alfn}
\end{equation}%
The eigenmodes of the dielectric cylinder are determined from the equation $%
\alpha _{n}=0$. They are poles of the integrand in (\ref{GFour}) for the
corresponding components of the Green tensor.

The Fourier components of the vector potential are given by (\ref{Anl}) with
the Green tensor components from (\ref{GFz}) where now $\omega =vk_{z}$ and $%
\lambda _{j}$ is given by the expression%
\begin{equation}
\lambda _{j}^{2}=k_{z}^{2}\left( \beta _{j}^{2}-1\right) ,  \label{lamj}
\end{equation}%
with $\beta _{j}^{2}=(v/c)^{2}\varepsilon _{j}$. In the discussion below we
will assume that the exterior medium is transparent \ and the permittivity $%
\varepsilon _{1}$ is real. The both cases $\beta _{1}>1$ and $\beta _{1}<1$
will be considered. In the second case $\lambda _{1}$ is purely imaginary
and its sign is determined in accordance with $\lambda _{1}=i|k_{z}|\sqrt{%
1-\beta _{1}^{2}}$. Note that in the arguments of the Hankel functions $%
\lambda _{1}$ appears only and with this choice of the sign they are reduced
to the Macdonald functions $K_{\nu }(|\lambda _{1}|x)$, with $%
x=r,r_{0},r_{c} $. For real $\varepsilon _{j}$ and $\beta _{j}^{2}>1$ the
signs are defined in accordance with $\lambda _{j}=k_{z}\sqrt{\beta
_{j}^{2}-1}$.

Fourier expanding the electric and magnetic fields $E_{l}(t,\mathbf{r})$ and
$H_{l}(t,\mathbf{r})$, similar to (\ref{vecF}), with the Fourier
coefficients $E_{l,n}(k_{z},r)$, $H_{l,n}(k_{z},r)$, for the magnetic field
one finds%
\begin{eqnarray}
H_{l,n}(k_{z},r) &=&\frac{qvk_{z}}{4i^{l-1}c}\sum_{p=\pm
1}p^{l-1}f_{n}^{(p)}H_{n+p}(\lambda _{1}r),\;l=1,2,  \notag \\
H_{3,n}(k_{z},r) &=&\frac{iqvk_{z}}{4c}\sqrt{\beta _{1}^{2}-1}\sum_{p=\pm
1}pf_{n}^{(p)}H_{n}(\lambda _{1}r),  \label{Hnz}
\end{eqnarray}%
where for $p=\pm 1$ we have defined the functions%
\begin{equation}
f_{n}^{(p)}=-\sqrt{\beta _{1}^{2}-1}J_{n}(\lambda _{1}r_{0})+\frac{%
H_{n}(\lambda _{1}r_{0})}{V_{n}^{H}}\left[ \sqrt{\beta _{1}^{2}-1}V_{n}^{J}+%
\frac{2ipk_{z}}{\pi }\frac{J_{n}(\lambda _{0}r_{c})}{r_{c}\alpha _{n}}\frac{%
J_{n+p}(\lambda _{0}r_{c})}{V_{n+p}^{H}}\right] .  \label{fnz}
\end{equation}%
By taking into account that for the function from (\ref{VF}) one has $%
V_{-n}^{F}=V_{n}^{F}$, $F=J,H$, it can be seen that
\begin{equation}
f_{-n}^{(p)}=(-1)^{n}f_{n}^{(-p)}.  \label{fnz1}
\end{equation}%
The Fourier coefficients for the electric field are obtained from the
Maxwell equations and are given by
\begin{eqnarray}
E_{l,n}(k_{z},r) &=&\frac{qk_{z}}{8i^{l}\varepsilon _{1}}\sum_{p=\pm 1}p^{l}%
\left[ \left( \beta _{1}^{2}+1\right) f_{n}^{(p)}-\left( \beta
_{1}^{2}-1\right) f_{n}^{(-p)}\right] H_{n+p}(\lambda _{1}r),  \notag \\
E_{3,n}(k_{z},r) &=&\frac{qk_{z}}{4\varepsilon _{1}}\sqrt{\beta _{1}^{2}-1}%
\sum_{p=\pm 1}f_{n}^{(p)}H_{n}(\lambda _{1}r),  \label{Enz}
\end{eqnarray}%
where $l=1,2$. From (\ref{fnz1}) we get the following relations for the
Fourier components of the fields%
\begin{equation}
E_{l,-n}(k_{z},r)=(-1)^{l+1}E_{l,n}(k_{z},r),%
\;H_{l,-n}(k_{z},r)=(-1)^{l}H_{l,n}(k_{z},r),  \label{relEH}
\end{equation}%
for $l=1,2,3$. Note that we have also the relations $%
E_{l,-n}(-k_{z},r)=E_{l,n}^{\ast }(k_{z},r)$ and $%
H_{l,-n}(-k_{z},r)=H_{l,n}^{\ast }(k_{z},r)$, where the star stands for the
complex conjugate.

The electromagnetic fields for a charge moving in a homogeneous medium with
dielectric permittivity $\varepsilon _{1}$ are obtained from the expressions
given above taking $\varepsilon _{0}=\varepsilon _{1}$. In this limit $%
V_{n}^{J}=0$ and $V_{n}^{H}=2i/\pi r_{c}$, whereas the function $\alpha _{n}$
tends to infinity. Hence, the corresponding Fourier components are given by (%
\ref{Hnz}) and (\ref{Enz}) with the replacement
\begin{equation}
f_{n}^{(p)}\rightarrow -\sqrt{\beta _{1}^{2}-1}J_{n}(\lambda _{1}r_{0}).
\label{ReplHom}
\end{equation}%
Now we see that the fields in the exterior region are decomposed into the
parts corresponding to the fields in homogeneous medium with permittivity $%
\varepsilon _{1}$ and the part induced by the presence of the cylinder. The
latter is given by (\ref{Hnz}) and (\ref{Enz}) excluding the first term in
the right-hand side of (\ref{fnz}). The Fourier components have poles at the
zeros of the function $\alpha _{n}$. As it has been mentioned before, those
zeros determine the eigenmodes of the cylinder. The expressions for the
fields in the region $r_{c}<r<r_{0}$ are obtained from the corresponding
formulas in the region $r>r_{0}$, given above, by the replacements $%
J\rightarrow H$, $H\rightarrow J$ in the parts corresponding to the fields
in homogeneous medium with permittivity $\varepsilon _{1}$. The cylinder
induced contributions are described by the same expressions for all values $%
r>r_{1}$. The fields inside the cylinder can be found by using the
corresponding expressions of the Green tensor components from \cite{Grig95}.

\section{Cherenkov radiation in the exterior medium}

\label{sec:Cher}

Having the electric and magnetic fields in the form of the Fourier expansion
we can investigate the radiation intensity emitted by the charged particle.
In the problem at hand we have three types of radiations. The first one
corresponds to the CR in the exterior medium influenced by the presence of
the cylinder. The second one is the radiation emitted on the guided modes of
the cylinder and propagates inside the waveguide. The corresponding fields
exponentially decay in the exterior medium. Under certain conditions on the
characteristics of the media one can have also the radiation in the form of
surface polaritons (surface modes). We start our discussion from the
radiation in the exterior medium at large distances from the cylinder, $r\gg
r_{c}$. From the expressions (\ref{Hnz}) and (\ref{Enz}) it follows that
this kind of radiation is present under the condition $\lambda _{1}^{2}>0$.
By taking into account the expression (\ref{lamj}) the latter condition is
translated to $\beta _{1}^{2}>1$ which is the Cherenkov condition for the
exterior medium. The corresponding radiation is the CR influenced by the
dielectric cylinder. For $\lambda _{1}^{2}<0$ the Hankel functions in (\ref%
{Hnz}) and (\ref{Enz}) are expressed in terms of the Macdonald functions $%
K_{n}(|\lambda _{1}|r)$, $K_{n+p}(|\lambda _{1}|r)$, and the Fourier
components exponentially decay at large distances from the cylinder, $r\gg
v/\omega $.

We denote by $I$ the energy flux per unit time through the cylindrical
surface of radius $r$. It is given by the expression
\begin{equation}
I=\frac{c}{4\pi }\int_{0}^{2\pi }d\phi \int_{-\infty }^{\infty }dz\,r\mathbf{%
n}\cdot \left[ \mathbf{E}\times \mathbf{H}\right] ,  \label{Ifl}
\end{equation}%
where $\mathbf{n}$ is the unit normal to the integration surface. By using
the Fourier expansions of the fields we get%
\begin{equation}
I=\pi cr\sum_{n=-\infty }^{\infty }\int_{-\infty }^{\infty }dk_{z}\,\mathbf{n%
}\cdot \left[ \mathbf{E}_{n}(k_{z},r)\times \mathbf{H}_{n}^{\ast }(k_{z},r)%
\right] .  \label{Ifl2}
\end{equation}%
Under the condition $\lambda _{1}^{2}>0$, substituting the expressions for
the Fourier components, using the relation (\ref{fnz1}) and the asymptotic
expressions of the Hankel functions for large arguments, at large distances
from the cylinder we find%
\begin{equation}
I=\int d\omega \,\frac{dI}{d\omega },  \label{Iint}
\end{equation}%
with the spectral density%
\begin{equation}
\frac{dI}{d\omega }=\frac{q^{2}\omega }{2v\varepsilon _{1}}%
\sideset{}{'}{\sum}_{n=0}^{\infty }\left[ \left\vert
f_{n}^{(1)}+f_{n}^{(-1)}\right\vert ^{2}+\beta _{1}^{2}\left\vert
f_{n}^{(1)}-f_{n}^{(-1)}\right\vert ^{2}\right] ,  \label{Ifl3}
\end{equation}%
where in the expressions (\ref{fnz}) for $f_{n}^{(\pm 1)}$ the quantities $%
\lambda _{j}$ are given by (\ref{lamj}) with $k_{z}=\omega /v$. In (\ref%
{Iint}), the integration over $\omega $ goes over the part of the region $%
\omega \in \lbrack 0,\infty )$ where the condition $\beta _{1}^{2}>1$ is
obeyed and the prime under the sign of the summation in (\ref{Ifl3}) means
that the term $n=0$ should be taken with an additional coefficient 1/2. An
alternative representation for the spectral distribution of the radiation
intensity $dI/d\omega $, based on the evaluation of the energy losses, will
be given below (see (\ref{Ifl3Loss})). In deriving (\ref{Ifl3}) we have used
the asymptotic expressions for the functions $H_{n}(\lambda _{1}r)$ and $%
H_{n+p}(\lambda _{1}r)$ in (\ref{Hnz}), (\ref{Enz}) valid in the range $%
\lambda _{1}r\gg 1$. This corresponds to the distances from the cylinder
axis much larger than the radiation wavelength. For a cylinder with finite
length $L_{c}$ additional conditions $r\ll L_{c}$ and $r_{c}\ll L_{c}$
should be imposed.

From the relation $\omega =k_{z}v$ it follows that the radiation described
by (\ref{Ifl3}) propagates along the Cherenkov cone having the opening angle
$\theta =\theta _{\mathrm{Ch}}$ with respect to the cylinder axis, where $%
\cos \theta _{\mathrm{Ch}}=1/\beta _{1}$. In the limit $\varepsilon
_{0}\rightarrow \varepsilon _{1}$ the functions $f_{n}^{(p)}$ are given by
the right-hand side in (\ref{ReplHom}) and from (\ref{Ifl3}) the Tamm-Frank
formula is obtained for the CR in a homogeneous transparent medium. In the
limit $r_{c}\rightarrow 0$ for fixed values of the other parameters, from (%
\ref{Ifl3}) to the leading order we obtain the corresponding result in a
homogeneous medium. The leading contribution to the cylinder induced part in
(\ref{Ifl3}) comes from the terms with $n=0,1$ and that contribution behaves
as $(\omega r_{c}/v)^{2}$. The contributions of the terms with $n\geq 2$
behave like $(\omega r_{c}/v)^{2n}$. Note that the quantity $\omega
^{-1}dI/d\omega $, that determines the number of the radiated quanta (see
below), depends on the frequency and on the cylinder radius in the form of
the product $\omega r_{c}$. Hence, the limiting behavior for small $r_{c}$
determines also the behaviour of the radiation intensity for small
frequencies. Namely, for $\omega r_{c}/v\ll 1$ the cylinder induced
contribution to the number of the radiated quanta behaves as $\omega ^{2}$
for the terms with $n=1,2$ and as $\omega ^{2n}$ for $n\geq 2$.

\begin{figure}[tbph]
\begin{center}
\begin{tabular}{cc}
\epsfig{figure=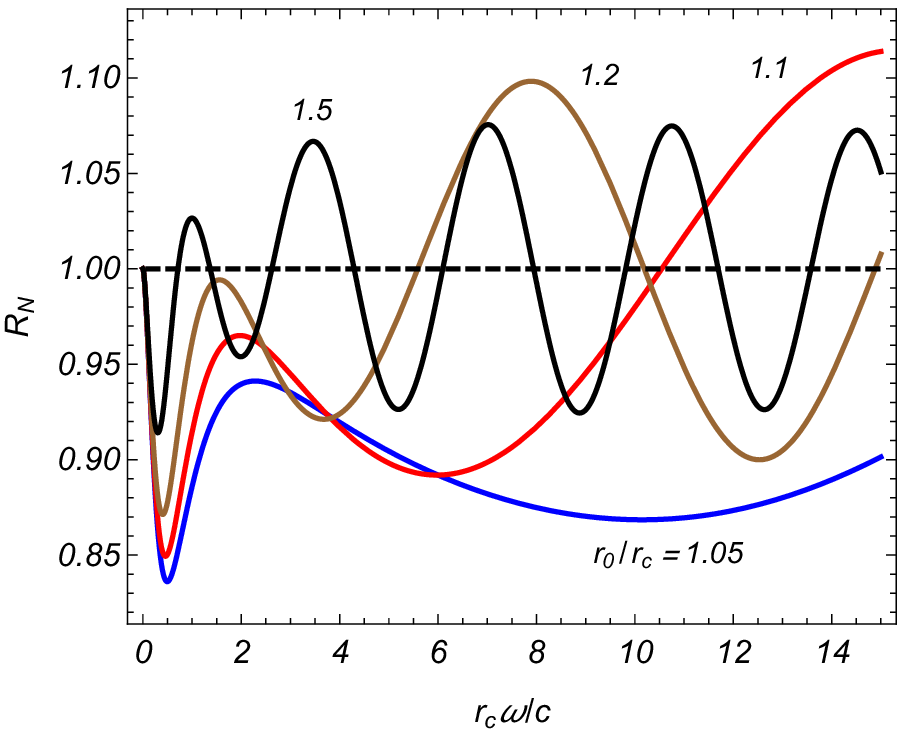,width=7.cm,height=5.5cm} & \quad %
\epsfig{figure=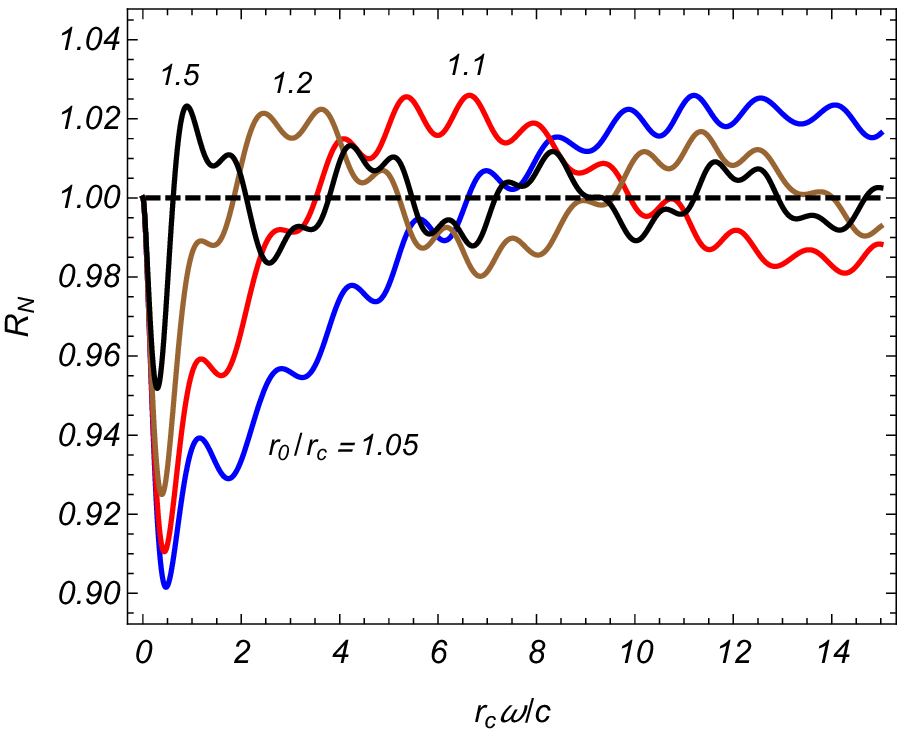,width=7.cm,height=5.5cm}%
\end{tabular}%
\end{center}
\caption{The ratio $R_{N}$ as a function of $\protect\omega r_{c}/c$ for the
electron energy $\mathcal{E}_{e}=2\,\mathrm{MeV}$ and for $\protect%
\varepsilon _{1}=3.8$. The left and right panels correspond to $\protect%
\varepsilon _{0}=1$ and $\protect\varepsilon _{0}=2.2$ and the numbers near
the curves are the values of $r_{0}/r_{c}$.}
\label{fig2}
\end{figure}

In figures below we present the spectral density of the number of photons
radiated per unit length of the charge trajectory:%
\begin{equation}
\frac{d^{2}N}{dzd\omega }=\frac{1}{\hbar \omega v}\frac{dI}{d\omega }.
\label{N}
\end{equation}%
The corresponding quantity for the CR in a transparent homogeneous medium
with permittivity $\varepsilon _{1}$ is given by%
\begin{equation}
\frac{d^{2}N_{0}}{dzd\omega }=\frac{q^{2}}{\hbar c^{2}}\left( 1-\frac{1}{%
\beta _{1}^{2}}\right) .  \label{N0}
\end{equation}%
In figure \ref{fig2} we display the ratio%
\begin{equation}
R_{N}=\frac{d^{2}N/dzd\omega }{d^{2}N_{0}/dzd\omega }  \label{RN}
\end{equation}%
as a function of $\omega r_{c}/c$ for several values of the ratio $%
r_{0}/r_{c}$ (the numbers near the curves). The graphs are plotted for the
electron energy $\mathcal{E}_{e}=2\,\mathrm{MeV}$ and for $\varepsilon
_{1}=3.8$ (average value for the real part of the dielectric permittivity
for fused quartz in the frequency range $\lesssim 1\,\mathrm{THz}$, in that
range the imaginary part of the permittivity is small, $\lesssim 10^{-3}$).
The left and right panels correspond to $\varepsilon _{0}=1$ and $%
\varepsilon _{0}=2.2$ (the real part of the dielectric permittivity for
teflon).

As seen from the graphs, we have characteristic oscillations with relatively
small shifts around the value corresponding to the radiation in a
homogeneous medium. The oscillation frequency increases with increasing $%
r_{0}/r_{c}$. In the case corresponding to the left panel of figure \ref%
{fig2} the CR inside the cylinder is absent and the oscillations are a
consequence of the interference between the direct CR and radiation
reflected from the cylinder. For small wavelengths, compared to the
waveguide diameter, the oscillations enter the quasiperiodic regime. The
beginning of that regime with respect to the radiation wavelength increases
with increasing values of the ratio $r_{0}/r_{c}$. For small frequencies the
presence of a cylindrical hole in a homogeneous medium leads to the decrease
of the radiation intensity. That is related to the fact that a part of the
medium is excluded from the radiation process. For the example considered on
the right panel of figure \ref{fig2} the Cherenkov condition for the
cylinder material is obeyed and the interference pattern is more
complicated. It is formed by the interference of the direct radiation, the
radiation reflected from the cylinder and the CR formed inside the cylinder.
We have $\varepsilon _{0}<\varepsilon _{1}$ and, as in the previous case,
here the radiation intensity for large wavelengths is smaller than that for
a homogeneous medium.

\begin{figure}[tbph]
\begin{center}
\begin{tabular}{cc}
\epsfig{figure=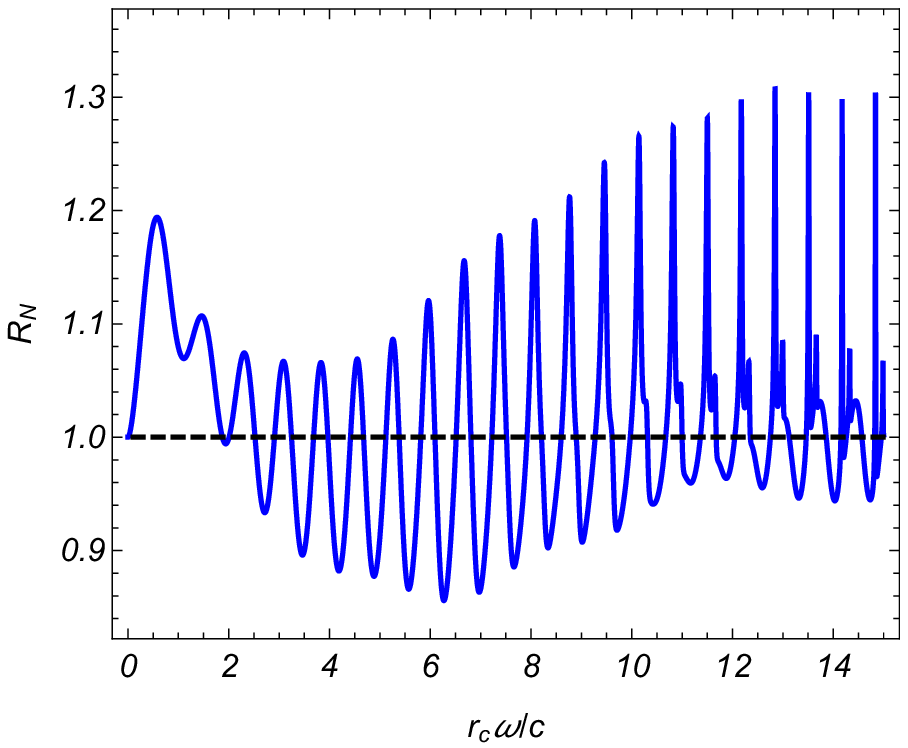,width=7.cm,height=5.5cm} & \quad %
\epsfig{figure=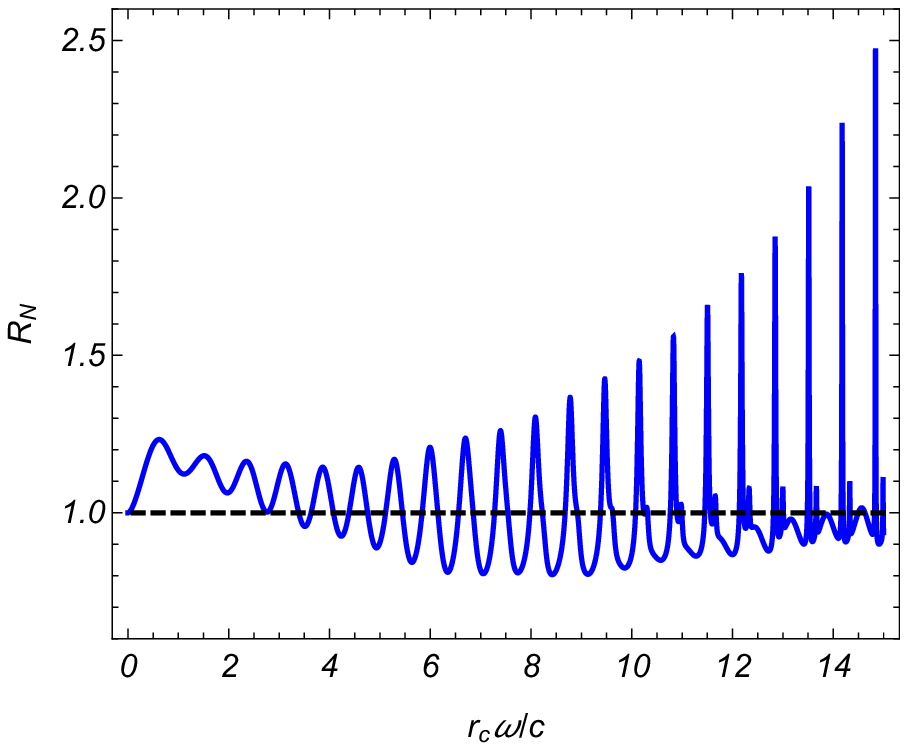,width=7.cm,height=5.5cm}%
\end{tabular}%
\end{center}
\caption{The same as in figure \protect\ref{fig2} for $\protect\varepsilon %
_{0}=3.8$, $\protect\varepsilon _{1}=2.2$. For the left and right panels $%
r_{0}/r_{c}=1.2$ and $r_{0}/r_{c}=1.1$.}
\label{fig3}
\end{figure}

\begin{figure}[tbph]
\begin{center}
\epsfig{figure=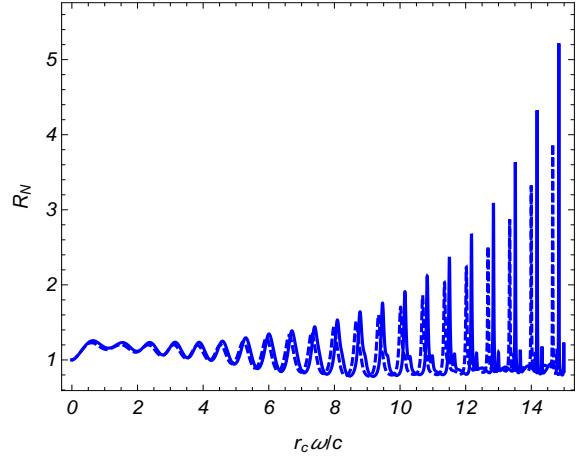,width=7.5cm,height=6cm}
\end{center}
\caption{The ratio $R_{N}$ versus $\protect\omega r_{c}/c$ for the electron
energies $\mathcal{E}_{e}=2\,\mathrm{MeV}$ (full curve) and $\mathcal{E}%
_{e}=10\,\mathrm{MeV}$ (dashed curve). The graphs are plotted for $%
r_{0}/r_{c}=1.05$, $\protect\varepsilon _{0}=3.8$, $\protect\varepsilon %
_{1}=2.2$. }
\label{fig4}
\end{figure}

For graphs in figure \ref{fig2} we have taken $\varepsilon _{1}>\varepsilon
_{0}$. The behavior of the radiation intensity is essentially different for $%
\varepsilon _{1}<\varepsilon _{0}$. This is seen from figures \ref{fig3} and %
\ref{fig4} where we have plotted $R_{N}$ versus $\omega r_{c}/c$ for $%
\varepsilon _{0}=3.8$, $\varepsilon _{1}=2.2$. In figure \ref{fig3} we have
taken $\mathcal{E}_{e}=2\,\mathrm{MeV}$, $r_{0}/r_{c}=1.2$ (left panel), $%
r_{0}/r_{c}=1.1$ (right panel). Figure \ref{fig4} is plotted for $%
r_{0}/r_{c}=1.05$ and for the energies $\mathcal{E}_{e}=2\,\mathrm{MeV}$
(full curve) and $\mathcal{E}_{e}=10\,\mathrm{MeV}$ (dashed curve). We have
numerically checked that the curves corresponding to the energies $\mathcal{E%
}_{e}>10\,\mathrm{MeV}$ practically coincide with those for the energy $10\,%
\mathrm{MeV}$. This is a consequence of the fact (also seen from the general
formula (\ref{Ifl3})) that the effects we consider are sensitive to the
velocity of the charge and not to the energy in the range $\mathcal{E}%
_{e}\gg m_{e}c^{2}$. As we see from the graphs, for the charge trajectory
sufficiently close to the cylinder strong narrow peaks appear in the
spectral density of the radiation intensity. The amplification of the
radiation intensity for relatively small values of $\omega r_{c}/c$ is
related to that now $\varepsilon _{0}>\varepsilon _{1}$ and the CR inside
the cylinder is more intense than in an equivalent cylinder with
permittivity $\varepsilon _{1}$.

The appearance of the strong narrow peaks in the spectral distribution of
the CR in the exterior medium is an interesting effect induced by the
cylinder. Their presence can be understood analytically by using the formula
(\ref{Ifl3}) for the radiation intensity (see also the discussions in \cite%
{Kota00} and \cite{Saha05} for the peaks in the angular distribution of the
radiation intensity from charges rotating around/inside a dielectric
cylinder along circular and helical trajectories, respectively). First of
all it can be seen that the peaks come from the terms in the series on the
right-hand side of (\ref{Ifl3}) with large values of $n$. For large $n$ one
has the following asymptotic expression for the Neumann function (the
leading term in the Debye's asymptotic expansion, see \cite{Abra72})%
\begin{equation}
Y_{n}(ny)\sim \frac{2e^{n\zeta (y)}}{\sqrt{2\pi n}(1-y^{2})^{1/4}},
\label{Yn}
\end{equation}%
with $0<y<1$ and%
\begin{equation}
\zeta (y)=\ln \frac{1+\sqrt{1-y^{2}}}{y}-\sqrt{1-y^{2}}.  \label{zeta}
\end{equation}%
The function (\ref{zeta}) is positive and monotonically decreasing in the
region $0<y<1$ with $\zeta (1)=0$. For $y>1$ and for large $n$ the function $%
Y_{n}(ny)$ exhibits an oscillating behavior (see the analog behavior for the
function $J_{n}(ny)$ in (\ref{Jn}) below). For the Bessel function one has
the asymptotics \cite{Abra72}%
\begin{eqnarray}
J_{n}(ny) &\sim &\frac{e^{-n\zeta (y)}}{\sqrt{2\pi n}(1-y^{2})^{1/4}}%
,\;0<y<1,  \notag \\
J_{n}(ny) &\sim &\sqrt{\frac{2}{\pi n}}\frac{\cos \{n[\sqrt{y^{2}-1}-\arccos
(1/y)]-\pi /4\}}{\left( y^{2}-1\right) ^{1/4}},\;y>1.  \label{Jn}
\end{eqnarray}%
The key point for our discussion is that the ratio $%
|J_{n}(ny_{1})|/Y_{n}(ny) $ is exponentially small for large $n$ and for
fixed $0<y<1$. For $0<y_{1}<1$ one has $J_{n}(ny_{1})/Y_{n}(ny)\propto
e^{-n[\zeta (y)+\zeta (y_{1})]}$ and for $y_{1}>1$ we get $%
|J_{n}(ny_{1})|/Y_{n}(ny)\propto e^{-n\zeta (y)}$.

With these asymptotic estimates, let us return to the expression (\ref{alfn}%
) for the function $\alpha _{n}$. As we have mentioned above, the roots of
the equation $\alpha _{n}=0$ determine the eigenmodes of the dielectric
cylinder. Under the condition $\lambda _{1}^{2}>0$ this equation has no
solutions. There are no eigenmodes in this range and all the radiated energy
goes to infinity in the form of the CR with the spectral density (\ref{Ifl3}%
). We can try to specify the conditions under which the function $\alpha
_{n} $ would take its minimal value. In accordance with (\ref{Ifl3}) that
could correspond to large intensities for the CR. By taking into account
that, in accordance with the asymptotics given above, for large $n$ and $%
\lambda _{1}r_{c}<n$ the ratio $|J_{n}(\lambda _{1}r_{c})|/Y_{n}(\lambda
_{1}r_{c})$ is exponentially small, we can expand $\alpha _{n}$ in terms of
this ratio. In the next-to-leading order we get
\begin{equation}
\alpha _{n}\approx \frac{\varepsilon _{0}}{\varepsilon _{1}-\varepsilon _{0}}%
+\frac{1}{2}\sum_{l=\pm 1}g_{l,n}+\frac{i}{\pi \lambda _{0}r_{c}}\sum_{l=\pm
1}\frac{lJ_{n+l}(\lambda _{0}r_{c})g_{l,n}^{2}}{J_{n}(\lambda
_{0}r_{c})Y_{n+l}^{2}(\lambda _{1}r_{c})},  \label{alfnexp}
\end{equation}%
where%
\begin{equation}
g_{l,n}=\left[ 1-\frac{\lambda _{1}}{\lambda _{0}}\frac{J_{n+l}(\lambda
_{0}r_{c})Y_{n}(\lambda _{1}r_{c})}{J_{n}(\lambda _{0}r_{c})Y_{n+l}(\lambda
_{1}r_{c})}\right] ^{-1}.  \label{gln}
\end{equation}%
Note that, compared to the first two terms in the right-hand side of (\ref%
{alfnexp}), the last term is of the order $e^{-2n\zeta (\lambda
_{1}r_{c}/n)} $. From here it follows that near the roots of the equation
\begin{equation}
\sum_{l=\pm 1}g_{l,n}+\frac{2\varepsilon _{0}}{\varepsilon _{1}-\varepsilon
_{0}}=0,  \label{Peaks}
\end{equation}%
the function $\alpha _{n}$ is exponentially small, $\alpha _{n}\propto
e^{-2n\zeta (\lambda _{1}r_{c}/n)}$. Of course, this does not yet mean that
the radiation intensity at those points will be large because exponential
factors may also come from the other functions in the last term of the
right-hand side of (\ref{fnz}).

We recall that under the condition $\lambda _{1}^{2}>0$ the equation $\alpha
_{n}=0$ has no solutions and there are no eigenmodes of the cylinder in that
region. The mathematical reason is that the function is complex and the real
and imaginary parts do not become zero simultaneously. Unlike the function $%
\alpha _{n}$, the function $g_{l,n}$ is real and the equation (\ref{Peaks})
may have solutions. In order to specify the conditions under which the roots
exist, first we consider the case $\lambda _{0}^{2}>0$ when the Cherenkov
condition for the material of cylinder is obeyed. For $\lambda _{0}r_{c}<n$,
by using the asymptotics (\ref{Yn}) and (\ref{Jn}) for the functions $%
Y_{n}(\lambda _{1}r_{c})$ and $J_{n}(\lambda _{0}r_{c})$, to the leading
order, the equation (\ref{Peaks}) is reduced to $\sqrt{n^{2}-\lambda
_{1}^{2}r_{c}^{2}}/\sqrt{n^{2}-\lambda _{0}^{2}r_{c}^{2}}=-\varepsilon
_{1}/\varepsilon _{0}$. This shows that for large values of $n$ and for $%
\lambda _{1}r_{c}<n$ the equation (\ref{Peaks}) has solutions under the
condition $\lambda _{0}r_{c}>n$. In particular, one should have $\varepsilon
_{0}>\varepsilon _{1}$. By making use of the uniform asymptotic expansion
for the modified Bessel function $I_{n}(|\lambda _{0}|r_{c})$, we can see
that from (\ref{Peaks}) the same leading order equation is obtained for $%
\lambda _{0}^{2}<0$. From that equation, as a necessary condition for the
existence of the roots in the range $\lambda _{0}^{2}<0<\lambda _{1}^{2}$
one gets $\varepsilon _{0}<-\varepsilon _{1}$. In the leading order, the
roots with respect to the angular frequency are given by%
\begin{equation}
\omega \approx \frac{cn}{r_{c}}\left( \frac{\varepsilon _{0}\varepsilon _{1}%
}{\varepsilon _{0}+\varepsilon _{1}}-\frac{c^{2}}{v^{2}}\right) ^{-1/2}.
\label{PeaksSP}
\end{equation}
Note that the inequality $\varepsilon _{0}<-\varepsilon _{1}$ appears also
as a necessary condition for the radiation of surface polaritons (see
below). For the latter modes one has $\lambda _{1}^{2}<0$ and they are
localized near the cylinder boundary.

Having specified the necessary conditions for the appearance of the peaks,
we can estimate the corresponding heights and widths. First of all, on the
base of the asymptotics for the Neumann and Bessel functions in the
expression (\ref{fnz}) of the functions $f_{n}^{(p)}$, it can be seen that
for the appearance of the peaks an additional condition $\lambda _{1}r_{0}<n$
is required. Under this condition, for the Hankel function in (\ref{fnz})
one has $H_{n}(\lambda _{1}r_{0})\approx iY_{n}(\lambda _{1}r_{0})$ and $%
f_{n}^{(p)}\propto e^{n\zeta (\lambda _{1}r_{0}/n)}$. As a consequence, the
heights of the peaks in the spectral distribution of the radiation intensity
are estimated as $e^{2n\zeta (\lambda _{1}r_{0}/n)}$. We have numerically
checked that the locations of the peaks with respect to $\omega r_{c}/c$ in
the graphs above are determined by the roots of the equation (\ref{Peaks})
with high accuracy. For example, the peaks in figure \ref{fig4} at $\omega
r_{c}/c=10.83,11.51,12.17,12.85,13.51,14.18$ come from the terms in (\ref%
{Ifl3}) with $n=15,16,17,18,19,20$, respectively. On the base of the
asymptotic consideration given above the widths of the peaks can be
estimated as well. In order to do that we expand the function $\alpha _{n}$
near the roots of the equation (\ref{Peaks}). By using (\ref{alfnexp}) it
can be seen that the width of the peaks is determined by the last term in
the right-hand side and is of the order $\Delta \omega /\omega \propto
e^{-2n\zeta (\lambda _{1}r_{c}/n)}$. Note that in the estimates given above
we have assumed that the dielectric permittivity $\varepsilon _{0}$ is real.
For complex permittivity $\varepsilon _{0}=\varepsilon _{0}^{\prime
}+i\varepsilon _{0}^{\prime \prime }$, with real and imaginary parts $%
\varepsilon _{0}^{\prime }$ and $\varepsilon _{0}^{\prime \prime }$, the
consideration presented is valid under the condition $e^{-2n\zeta (\lambda
_{1}r_{c}/n)}\gg |\varepsilon _{0}^{\prime \prime }/\varepsilon _{0}^{\prime
}|$. For $e^{-2n\zeta (\lambda _{1}r_{c}/n)}<$ $|\varepsilon _{0}^{\prime
\prime }/\varepsilon _{0}^{\prime }|$ the heights and the widths of the
peaks are determined by the imaginary part of the permittivity.

Summarizing the discussion above, we conclude that though there are no
eigenmodes of the waveguide in the range under consideration ($\lambda
_{1}^{2}>0$), the equation (\ref{Peaks}) may have roots and for large values
of $n$ they approximately obey the equation $\alpha _{n}=0$ with exponential
accuracy. In this sense, those roots can be termed as "quasimodes" of the
dielectric waveguide (for the discussion of quasi-bound waves on curved
interfaces see, e.g., \cite{Berr75}). Unlike to the guided and surface modes
(see below) which remain coupled to the waveguide during their propagation
and exponentially decay in the exterior medium, the radiation on the
"quasimodes" appears in form of the CR giving rise to high narrow peaks in
the spectral distribution of the radiation intensity under the conditions $%
\lambda _{1}r_{0}<n<\lambda _{0}r_{c}$ for $\lambda _{0}^{2}>0$ and under
the conditions $\lambda _{1}r_{0}<n$, $\varepsilon _{0}<-\varepsilon _{1}$
for $\lambda _{0}^{2}<0$. In the latter case, for a given $n$, the angular
frequencies of the peaks are given by (\ref{PeaksSP}). In the corresponding
spectral range one has a quasidiscrete part of the CR. The spectral peaks
appear for large values of $n$ and, hence, this effect is absent in the
axially symmetric problems (coaxial motion of charges and beams) where only
the mode $n=0$ contributes to the radiation intensity. We expect that
similar features of the radiation intensity may appear for other geometries
of the interface (see, for example, Ref. \cite{Grig06} for the radiation on
a dielectric ball).

In the consideration above, in order to have an exactly solvable problem, we
have made a number of idealizations. The possibility of experimental
observation of the features discussed requires further investigations by
taking into account a number of additional factors that can affect the
radiation characteristics. In particular, they include the finite thickness
of the medium where the particle moves (the exterior medium in the problem
under consideration), the finite length of the waveguide, the collective
effects of the particles in the bunch when the bunch size is of the order of
radiation wavelength or larger, the shift of the particle trajectory from
the one we have considered. Similar to the case of the standard Cherenkov
radiation in dielectric plates, the finite thickness of the radiator will
lead to broadening of the angular distribution of the radiation. We also
expect broadening of the peaks in the spectral distribution.

The charge moving in a medium suffers multiple scattering and this restricts
the mean length of straight trajectory. The multiple scattering leads to
beam broadening that is determined by root mean square (rms) scattering
angle $\theta _{\mathrm{ms}}$. The influence of beam broadening on the
Cherenkov radiation in a homogeneous medium has been investigated in the
literature both theoretically and experimentally (see, e.g., \cite{Dedr52}
and references therein). For small angles $\theta _{\mathrm{ms}}$, the beam
broadening leads to an additional factor in the angular-frequency
distribution of the radiated energy. Note that at relatively small energies
the multiple scattering may essentially restrict the length of the particle
straight trajectory in a medium. For example, for an electron with energy $%
2\,\mathrm{MeV}$, that we have taken above for illustrative purposes only,
the lengths in quartz and teflon are of the order of $1\,\mathrm{mm}$. With
increasing energy the scattering angle $\theta _{\mathrm{ms}}$ decreases
inversely proportional to the energy (see, for example, \cite{Zyla20}) and
the mean length of the straight trajectory increases. As it already has been
mentioned before, the features of the Cherenkov radiation we have discussed
are sensitive to the velocity of the particle and are not sensitive to the
particle energy at relatively high energies. For example, for the values of
the parameters corresponding to figure \ref{fig4} the locations of the peaks
and the corresponding heights are almost the same for all the energies
larger than $10\,\mathrm{MeV}$. Note that, in general, the thickness of the
exterior medium and the length of the waveguide can be different. The charge
moves in the exterior medium and the multiple scattering restricts the first
parameter only.

An interesting possibility to escape multiple scattering was indicated in
\cite{Bolo57,Ginz89}. It has been argued that an empty channel along the
particle trajectory in solid dielectric does not affect the radiation
intensity if the channel radius is smaller than the wavelength of the
radiation. The Cherenkov radiation by an electron bunch moving in a hollow
cylindrical channel in dielectric-lined waveguides has been experimentally
observed in \cite{Cook09} for the electron energies $\mathcal{E}_{e}=10\,%
\mathrm{MeV},\,60\,\mathrm{MeV}$ and for the radii of the channel $%
r_{c}=0.25\,\mathrm{mm},\,0.1\,\mathrm{mm}$, respectively. Hollow capillary
tubes with dielectric walls are among the main elements in dielectric
wakefield accelerators and in capillary-guided laser wakefield accelerators
(see \cite{Esar09,Thom08,Shea16} and references therein). Such schemes
provide relatively compact accelerating systems with large acceleration
gradients. In related experiments the parameters of the electron bunch and
the radius of the tube vary over wide ranges. For example, in \cite{Thom08}
the experiments were performed for the beam energy $28.5\,\mathrm{GeV}$, rms
bunch radius $0.01\,\mathrm{mm}$, rms bunch lengths from $0.01\,\mathrm{mm}$
to $0.1\,\mathrm{mm}$, and for the tube inner diameter $0.1\,\mathrm{mm}$.
The corresponding parameters for the experiments described in \cite{Shea16}
are given as $20.35\,\mathrm{GeV}$, $0.03\,\mathrm{mm}$, $0.025\,\mathrm{mm\,%
}-0.05\,\mathrm{mm}$, $\,$and $0.3\,\mathrm{mm}$. In both cases SiO$_{2}$
annular capillaries have been used. In \cite{Shea16} the length of the
cappilaries ranges from $1\,\mathrm{cm}$ to $15\,\mathrm{cm}$. In our setup,
a hollow cylinder along the particle trajectory, corresponding to the inner
region of this kind of cappilaries, will not influence the features of the
Cherenkov radiation in the frequency range $\lesssim 1\,\mathrm{THz}$.

\section{Energy losses}

\label{sec:Losses}

In addition to the radiation propagating at large distances from the
cylinder one can have radiation emitted by the charge on the eigenmodes of
the cylindrical waveguide. The total energy losses per unit of path length
can be evaluated in terms of the work done by the electromagnetic field on
the charge:%
\begin{equation}
\frac{dW}{dz}=qE_{3}|_{r\rightarrow r_{0},z\rightarrow vt,\phi \rightarrow
\phi _{0}}.  \label{W1}
\end{equation}%
Substituting the analog of the Fourier expansion (\ref{vecF}) for the $z$%
-component of the electric field one gets%
\begin{equation}
\frac{dW}{dz}=4q\sideset{}{'}{\sum}_{n=0}^{\infty }\mathrm{Re}\left[
\int_{0}^{\infty }dk_{z}\,E_{n3}(k_{z},r_{0})\right] ,  \label{W1b}
\end{equation}%
where the relation (\ref{relEH}) is used for $E_{n3}$. By using the
expression for the corresponding Fourier component (\ref{Enz}), we obtain%
\begin{equation}
\frac{dW}{dz}=q^{2}\sideset{}{'}{\sum}_{n=0}^{\infty }\mathrm{Re}\left[
\int_{0}^{\infty }dk_{z}\,\frac{k_{z}}{\varepsilon _{1}}\sqrt{\beta
_{1}^{2}-1}\sum_{p=\pm 1}f_{n}^{(p)}H_{n}(\lambda _{1}r)\right] .  \label{W2}
\end{equation}%
Note that, unlike the expression (\ref{Ifl3}), the functions $f_{n}^{(p)}$
enter in the expression of the energy losses linearly.

By taking into account the formulas (\ref{Enz}) and (\ref{fnz}), the
expression ( \ref{W1b}) is decomposed into two contributions:%
\begin{equation}
\frac{dW}{dz}=\frac{dW^{(0)}}{dz}+\frac{dW^{\mathrm{(c)}}}{dz},  \label{Wdec}
\end{equation}%
where
\begin{equation}
\frac{dW^{(0)}}{dz}=-2q^{2}\lim_{r\rightarrow r_{0}}\sideset{}{'}{\sum}%
_{n=0}^{\infty }\mathrm{Re}\left[ \int_{0}^{\infty }dk_{z}\,\frac{k_{z}}{%
\varepsilon _{1}}\left( \beta _{1}^{2}-1\right) J_{n}(\lambda
_{1}r_{0})H_{n}(\lambda _{1}r)\right] ,  \label{W0}
\end{equation}%
corresponds to the losses in a homogeneous medium with permittivity $%
\varepsilon _{1}$ and
\begin{eqnarray}
\frac{dW^{\mathrm{(c)}}}{dz} &=&2q^{2}\sideset{}{'}{\sum}_{n=0}^{\infty }\,%
\mathrm{Re}\left\{ \int_{0}^{\infty }dk_{z}\frac{k_{z}}{\varepsilon _{1}}%
\left( \beta _{1}^{2}-1\right) \frac{H_{n}^{2}(\lambda _{1}r_{0})}{V_{n}^{H}}%
\right.  \notag \\
&&\times \left. \left[ V_{n}^{J}+\frac{ik_{z}J_{n}(\lambda _{0}r_{c})}{\pi
\sqrt{\beta _{1}^{2}-1}r_{c}\alpha _{n}}\sum_{p=\pm 1}p\frac{J_{n+p}(\lambda
_{0}r_{c})}{V_{n+p}^{H}}\right] \right\}  \label{Wc}
\end{eqnarray}%
is induced by the cylinder. This formula gives the expression for the losses
in the general case of the dielectric permittivity for the cylinder.

First we consider the case when the Cherenkov condition for the exterior
medium is satisfied, $\beta _{1}^{2}>1$. The part (\ref{W0}) is further
simplified by using the formula%
\begin{equation}
\sideset{}{'}{\sum}_{n=0}^{\infty }J_{n}(\lambda _{1}r_{0})H_{n}(\lambda
_{1}r)=\frac{1}{2}H_{0}\left( \lambda _{1}(r-r_{0})\right) .  \label{sumn}
\end{equation}%
The real part of the latter is $J_{0}\left( \lambda _{1}(r-r_{0})\right) /2$
and taking the limit $r\rightarrow r_{0}$ in (\ref{W0}) we get%
\begin{equation}
\frac{dW^{(0)}}{dz}=-\frac{q^{2}}{c^{2}}\int_{\beta _{1}>1}d\omega \,\omega
\left( 1-1/\beta _{1}^{2}\right) ,  \label{W0b}
\end{equation}%
which gives the standard expression for the Cherenkov radiation in
homogeneous medium. Under the condition $\beta _{1}^{2}>1$ one has $\lambda
_{1}^{2}>0$ and it can be shown that the equation $\alpha _{n}=0$ has no
solutions with respect to $k_{z}$ and the integrand in (\ref{Wc}) is regular
on the positive semiaxis of $k_{z}$. For real values of $\varepsilon
_{0}=\varepsilon _{0}(\omega )$ the energy losses are in the form of the
radiation (here and below we will not consider the ionization losses that
correspond to the zeros of the function $\varepsilon _{1}$). For the
spectral density of the energy radiated per unit time we find%
\begin{eqnarray}
\frac{dI}{d\omega } &=&-v\frac{d^{2}W}{dzd\omega }=\frac{q^{2}v}{c^{2}}%
\,\omega \left( 1-\frac{1}{\beta _{1}^{2}}\right) \left\{ 1-2%
\sideset{}{'}{\sum}_{n=0}^{\infty }\,\mathrm{Re}\left[ \frac{%
H_{n}^{2}(\lambda _{1}r_{0})}{V_{n}^{H}}\right. \right.  \notag \\
&&\times \left. \left. \left( V_{n}^{J}+\frac{ik_{z}J_{n}(\lambda _{0}r_{c})%
}{\pi \sqrt{\beta _{1}^{2}-1}r_{c}\alpha _{n}}\sum_{p=\pm 1}p\frac{%
J_{n+p}(\lambda _{0}r_{c})}{V_{n+p}^{H}}\right) \right] \right\} .
\label{Ifl3Loss}
\end{eqnarray}%
For $\beta _{1}^{2}>1$ there is no radiation on the eigenmodes of the
cylinder and (\ref{Ifl3Loss}) corresponds to the CR in the exterior medium.
We have numerically checked that the formula (\ref{Ifl3Loss}) gives the same
results as (\ref{Ifl3}) for both the cases $\lambda _{0}^{2}>0$ and $\lambda
_{0}^{2}<0$. Note that in (\ref{Ifl3Loss}) the contribution corresponding to
the radiation in homogeneous medium (the part with the first term in figure
braces) is explicitly separated.

\section{Radiation on guided modes of the waveguide}

\label{sec:Guiding}

Now we consider the case when the Cherenkov condition for the exterior
medium is not obeyed, $\beta _{1}<1$. In this case one has $\lambda
_{1}=i|\lambda _{1}|$ and the expression in the square brackets of (\ref{W0}%
) is purely imaginary. As a consequence, we get $dW^{(0)}/dz=0$ and the
radiation in a transparent homogeneous medium is absent. Introducing the
modified Bessel functions $I_{n}(x)$ and $K_{n}(x)$, the expression for the
energy losses is presented as
\begin{eqnarray}
\frac{dW}{dz} &=&-\frac{2q^{2}}{\pi }\sideset{}{'}{\sum}_{n=0}^{\infty }%
\mathrm{Im}\,\left\{ \int_{0}^{\infty }dk_{z}\frac{k_{z}}{\varepsilon _{1}}%
\left( 1-\beta _{1}^{2}\right) \frac{K_{n}^{2}(|\lambda _{1}|r_{0})}{%
V_{n}^{J,K}}\right.  \notag \\
&&\times \left. \left[ 2V_{n}^{J,I}-\frac{k_{z}J_{n}(\lambda _{0}r_{c})}{%
r_{c}\sqrt{1-\beta _{1}^{2}}\alpha _{n}}\sum_{p=\pm 1}\frac{J_{n+p}(\lambda
_{0}r_{c})}{V_{n+p}^{J,K}}\right] \right\} .  \label{Wc3}
\end{eqnarray}%
with ($F=I,K$)%
\begin{equation}
V_{n}^{J,F}=J_{n}(\lambda _{0}r_{c})\partial _{r_{c}}F_{n}(|\lambda
_{1}|r_{c})-F_{n}(|\lambda _{1}|r_{c})\partial _{r_{c}}J_{n}(\lambda
_{0}r_{c}),  \label{VJF}
\end{equation}%
and
\begin{equation}
\alpha _{n}=\frac{\varepsilon _{0}}{\varepsilon _{1}-\varepsilon _{0}}+\frac{%
1}{2}\sum_{l=\pm 1}\left[ 1+l\frac{|\lambda _{1}|}{\lambda _{0}}\frac{%
J_{n+l}(\lambda _{0}r_{c})K_{n}(|\lambda _{1}|r_{c})}{J_{n}(\lambda
_{0}r_{c})K_{n+l}(|\lambda _{1}|r_{c})}\right] ^{-1}.  \label{alfn3}
\end{equation}%
For real values of $\varepsilon _{0}$ the integrand in (\ref{Wc3}) is real
on the real axis of $k_{z}$ and the nonzero contributions to the integral
may come from the possible poles on the real axis only. We can see that the
integral is regular at the zeros of the functions $V_{n\pm 1}^{J,K}$ and $%
V_{n}^{J,K}$. Hence, the only nonzero contributions come from the zeros of
the function $\alpha _{n}$. These zeros with respect to $k_{z}$ we will
denote by $k_{z}=k_{n,s}>0$, where $s=1,2,\ldots $ enumerates the roots for
a given $n$, $k_{n,s+1}>k_{n,s}$. These roots determine the eigenmodes of
the dielectric cylinder (the equation $\alpha _{n}=0$ is easily transformed
to the form given, for example, in \cite{Jack99}). For $n=0$ the equation
for those modes is simplified to
\begin{equation}
\varepsilon _{0}\frac{\sqrt{1-\beta _{1}^{2}}}{\sqrt{\beta _{0}^{2}-1}}\frac{%
J_{1}(\lambda _{0}r_{c})}{J_{0}(\lambda _{0}r_{c})}+\varepsilon _{1}\frac{%
K_{1}(|\lambda _{1}|r_{c})}{K_{0}(|\lambda _{1}|r_{c})}=0.  \label{n0mode}
\end{equation}%
Note that the product $k_{z}r_{c}=k_{n,s}r_{c}$ does not depend on $r_{c}$
and is a function of two parameters, $\varepsilon _{0}/\varepsilon _{1}$ and
$\beta _{1}$,%
\begin{equation}
k_{n,s}r_{c}=f(\varepsilon _{0}/\varepsilon _{1},\beta _{1}).  \label{knsf}
\end{equation}

In order to evaluate the integral in (\ref{Wc3}) one needs to specify the
integration contour near the poles $k_{z}=k_{n,s}$. In this section we will
consider the spectral range where $\lambda _{0}^{2}>0$. The corresponding
eigenmodes $k_{n,s}$ are the guided modes of the dielectric cylinder. For
those modes the radial dependence of the Fourier components for the fields
inside the cylinder is expressed in terms of the Bessel function $%
J_{n}(\lambda _{0}r)$. In order to specify the contour, we note that in
physically realistic problems the permittivity $\varepsilon _{0}$ has an
imaginary part, $\varepsilon _{0}=\varepsilon _{0}^{\prime }+i\varepsilon
_{0}^{\prime \prime }$. We consider $\alpha _{n}$ from (\ref{alfn3}) as a
function of $k_{z}$ and $\varepsilon _{0}$, $\alpha _{n}=\alpha
_{n}(k_{z},\varepsilon _{0})$. Note that in the presence of dispersion one
has $\varepsilon _{0}=\varepsilon _{0}(\omega )=\varepsilon _{0}(k_{z}v)$
and the second argument is a function of $k_{z}$ as well. Assuming that $%
|\varepsilon _{0}^{\prime \prime }/\varepsilon _{0}^{\prime }|\ll 1$, the
dominant contribution to the integral in (\ref{Wc3}) comes from the region
near $k_{z}=k_{n,s}$, where $k_{n,s}$ is the $s$th root of the equation $%
\alpha _{n}(k_{z},\varepsilon _{0}^{\prime })=0$. First we write $\alpha
_{n}(k_{z},\varepsilon _{0})\approx \alpha _{n}(k_{z},\varepsilon
_{0}^{\prime })+i\varepsilon _{0}^{\prime \prime }\partial _{\varepsilon
_{0}^{\prime }}\alpha _{n}(k_{z},\varepsilon _{0}^{\prime })$ and then
expand near $k_{z}=k_{n,s}$:%
\begin{equation}
\alpha _{n}(k_{z},\varepsilon _{0})\approx \partial _{k_{z}}\alpha
_{n}(k_{z},\varepsilon _{0}^{\prime })|_{k_{z}=k_{n,s}}\left(
k_{z}-k_{m,s}+i\varepsilon _{0}^{\prime \prime }b_{n,s}\right) ,
\label{alfExp}
\end{equation}%
where
\begin{equation}
b_{n,s}=\left. \frac{\partial _{\varepsilon _{0}}\alpha
_{n}(k_{z},\varepsilon _{0})}{\partial _{k_{z}}\alpha _{n}(k_{z},\varepsilon
_{0})}\right\vert _{k_{z}=k_{n,s},\varepsilon _{0}=\varepsilon _{0}^{\prime
}}.  \label{bns}
\end{equation}%
Note that, though $\varepsilon _{0}$ may depend on $k_{z}$, the derivative $%
\partial _{\varepsilon _{0}}\alpha _{n}(k_{z},\varepsilon _{0})$ is taken
for the fixed value of $k_{z}$. As for the denominator, $\partial
_{k_{z}}\alpha _{n}(k_{z},\varepsilon _{0})=(d/dk_{z})\alpha
_{n}(k_{z},\varepsilon _{0})$, in the presence of dispersion $\varepsilon
_{0}=\varepsilon _{0}(k_{z}v)$ the derivative is taken with respect to both
the arguments. From (\ref{alfExp}) we see that the pole of the integrand in (%
\ref{Wc3}) is located at $k_{z}=k_{m,s}-i\varepsilon _{0}^{\prime \prime
}b_{n,s}$. We have numerically checked that the numerator in (\ref{bns}) is
negative for $\lambda _{0}^{2}>0$ and the sign of $b_{n,s}$ is determined by
the sign of the denominator. The latter will be denoted as $\sigma _{n,s}=%
\mathrm{sgn}(\partial _{k_{z}}\alpha _{n}(k_{z},\varepsilon _{0}^{\prime
})|_{k=k_{n,s}})=-\mathrm{sgn}(b_{n,s})$. By taking into account that $%
\varepsilon _{0}^{\prime \prime }(\omega )>0$ for $\omega >0$, from here we
conclude that for $\lambda _{0}^{2}>0$ in (\ref{Wc3}) the poles $%
k_{z}=k_{n,s}$ should be avoided from above for $\sigma _{n,s}<0$ and from
below for $\sigma _{n,s}>0$ by small semicircles in the complex plane $k_{z}$%
. The integrals over these semicircles are expressed in terms of the
corresponding residues. Returning to the case of real $\varepsilon _{0}$, $%
\varepsilon _{0}=\varepsilon _{0}^{\prime }$, for the energy radiated per
unit time we get%
\begin{equation}
I=\sideset{}{'}{\sum}_{n=0}^{\infty }\sum_{s}I_{n,s}=-v\frac{dW}{dz},
\label{Ins}
\end{equation}%
where the radiation intensity on the angular frequency $\omega
_{n,s}=vk_{n,s}$ is given by
\begin{equation}
I_{n,s}=-2\delta _{n}q^{2}\frac{v}{\varepsilon _{1}}\sqrt{1-\beta _{1}^{2}}%
\left. k_{z}^{2}\frac{K_{n}^{2}(|\lambda _{1}|r_{0})}{V_{n}^{J,K}}\frac{%
J_{n}(\lambda _{0}r_{c})}{r_{c}|\alpha _{n}^{\prime }(k_{z})|}\sum_{p=\pm 1}%
\frac{J_{n+p}(\lambda _{0}r_{c})}{V_{n+p}^{J,K}}\right\vert _{k_{z}=k_{n,s}}.
\label{Wc3b}
\end{equation}%
Here, $\alpha _{n}^{\prime }(k_{z})=\partial _{k_{z}}\alpha
_{n}(k_{z},\varepsilon _{0})$, $\delta _{0}=1/2$ and $\delta _{n}=1$ for $%
n=1,2,\ldots $. This expression determines the radiation intensity on the
guided modes of the dielectric waveguide. If the Cherenkov condition for the
surrounding medium is not satisfied the CR emitted inside the cylinder is
totally reflected from the separating boundary.

The dependence of the radiation intensity on the distance of the charge from
the waveguide axis enters through the function $K_{n}^{2}(|\lambda
_{1}|r_{0})$. For large values of $r_{0}$ the intensity is exponentially
small. For large values of $|\lambda _{1}|r_{c}\gg 1$, the intensity is
suppressed by the factor $e^{-2|\lambda _{1}|(r_{0}-r_{c})}$. Hence, the
guided modes of the waveguide are mainly radiated on the frequencies%
\begin{equation}
\omega _{n,s}\lesssim \frac{v}{\sqrt{1-\beta _{1}^{2}}(r_{0}-r_{c})}.
\label{omnlim}
\end{equation}%
For $n\geq 1$ one has $k_{n+1,1}>k_{n,1}$. In table \ref{tab1} we present $%
k_{n,1}r_{c}$ for $\mathcal{E}_{e}=2\,\mathrm{MeV}$, $\varepsilon _{0}=3.8$,
$\varepsilon _{1}=1$ and for several values of $n$. As seen, for $n\geq 1$
the first root $k_{n,1}r_{c}$ is of the order of $n$.

\begin{table}[tbp]
\caption{The first eigenvalues for $k_{z}r_{c}$ for different values of the
azimuthal number $n$.}
\label{tab1}\centering
\begin{tabular}{|c|c|c|c|c|c|c|c|c|c|}
\hline\hline
$n$ & 0 & 1 & 2 & 3 & 4 & 5 & 10 & 15 & 20 \\ \hline
$k_{n,1}r_{c}$ & 1.689 & 0.886 & 1.971 & 2.866 & 3.685 & 4.465 & 8.124 &
11.613 & 15.027 \\ \hline
\end{tabular}%
\end{table}
Assuming that $|\lambda _{j}|r_{c}\gg n$, the asymptotic expression for the
roots $k_{n,s}$ is found by using the asymptotic formulas for the cylinder
functions for large arguments:%
\begin{equation}
k_{n,2l+1}r_{c}\approx \frac{1}{\sqrt{\beta _{0}^{2}-1}}\left[ \frac{n\pi }{2%
}+\frac{\pi }{4}-\arctan \left( \frac{\varepsilon _{1}}{\varepsilon _{0}}%
\sqrt{\frac{\beta _{0}^{2}-1}{1-\beta _{1}^{2}}}\right) +\pi l\right] ,
\label{kas}
\end{equation}%
where $l\gg 1$ and $k_{n,2l}<k_{n,2l+1}$ is close to (\ref{kas}). For a
given $n$, the frequency $\omega _{n,s}$ of the guided mode increases with
increasing $s$ and the upper limit of the summation over $s$ in (\ref{Ins})
is determined from the Cherenkov condition $v\varepsilon _{0}(\omega
_{n,s})/c>1$.

\begin{figure}[tbph]
\begin{center}
\begin{tabular}{cc}
\epsfig{figure=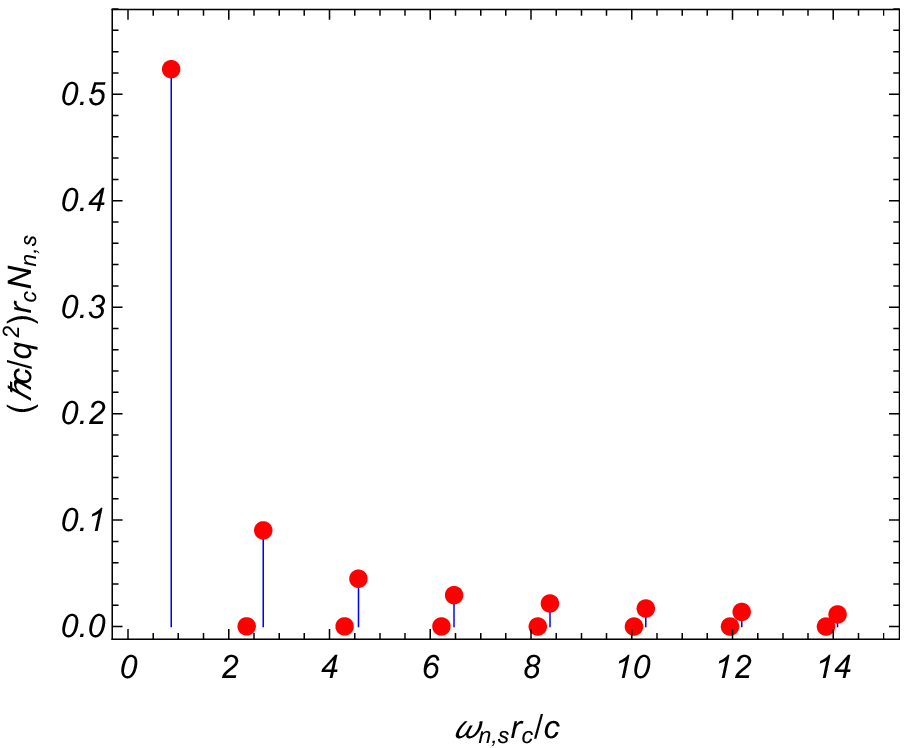,width=7.cm,height=5.5cm} & \quad %
\epsfig{figure=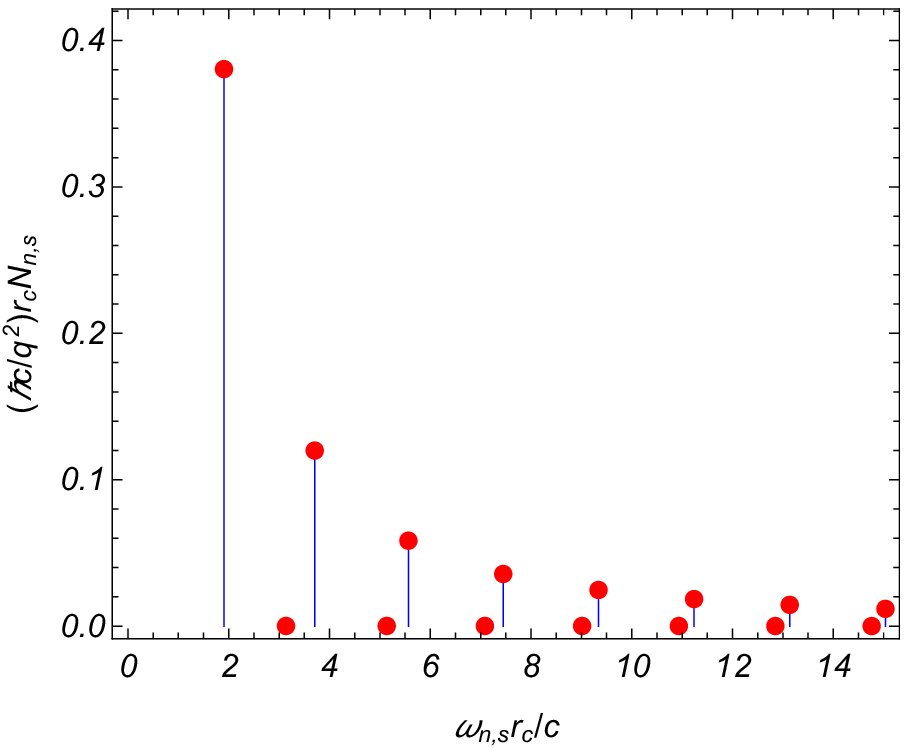,width=7.cm,height=5.5cm}%
\end{tabular}%
\end{center}
\caption{The number of quanta radiated on guided modes of the cylinder
versus $\protect\omega _{n,s}r_{c}/c$ for $n=1$ (left panel) and $n=2$
(right panel). The data are presented for $\mathcal{E}_{e}=2\,\mathrm{MeV}$,
$\protect\varepsilon _{0}=3.8$, $\protect\varepsilon _{1}=1$, $%
r_{0}/r_{c}=1.05$.}
\label{fig5}
\end{figure}

\begin{figure}[tbph]
\begin{center}
\epsfig{figure=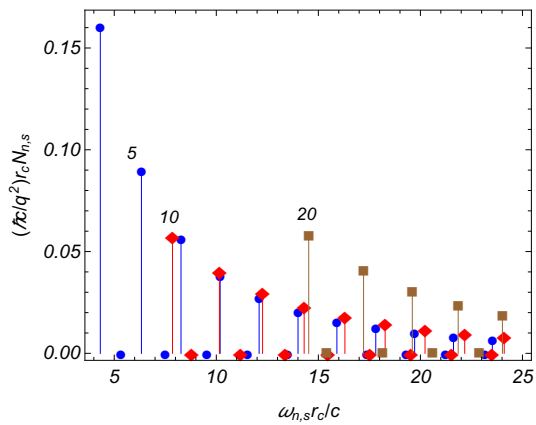,width=7.cm,height=5.5cm}
\end{center}
\caption{The same as in figure \protect\ref{fig5} for $n=5,10,20$.}
\label{fig6}
\end{figure}

In figures below we plot the number of quanta radiated on a given mode $%
k_{n,s}$ per unit length of the charge trajectory:%
\begin{equation}
N_{n,s}=\frac{I_{n,s}}{\hbar \omega _{n,s}v}.  \label{Nns}
\end{equation}%
Figure \ref{fig5} presents the number of the radiated quanta as a function
of $\omega _{n,s}r_{c}/c$ for given $n$ and for different values of $s$. For
the parameters we have taken $\mathcal{E}_{e}=2\,\mathrm{MeV}$, $\varepsilon
_{0}=3.8$, $\varepsilon _{1}=1$, $r_{0}/r_{c}=1.05$. The left and right
panels correspond to $n=1$ and $n=2$, respectively. As seen, for fixed $n$
and started from $s=2$ the roots $k_{n,s}$ come in pairs which are close to
each other. The radiation intensity on the first root in the pair is much
smaller than on the second one. For example $N_{1,2}/N_{1,3}\approx 0.0026$
and $N_{2,2}/N_{2,3}\approx 0.001$. The radiation on the modes with $n=0$ is
essentially smaller compared to the cases presented in figure \ref{fig5}.
For the same values of the parameters one has $\omega _{0,1}r_{c}/c\approx
1.63$ and $r_{c}N_{0,1}\approx 0.026q^{2}/(\hbar c)$. The corresponding
results for $n=5$ (circles), $n=10$ (diamonds) and $n=20$ (squares) are
presented in figure \ref{fig6}.

Note that the numerical results above are given in relative units and can be
used to estimate the radiation intensity in wide range of frequencies. The
absolute values for the radiation frequencies depend on the diameter of the
cylindrical waveguide and are restricted by the condition (\ref{omnlim}).
For available waveguides the diameter may vary over a wide range started
from 50 nm (used for optical wave guiding). In particular, various types of
terahertz waveguides, with radius of the order 1 mm, have been discussed in
the literature. Note that in figures given above (and also given below for
the radiation of surface polaritons) we have plotted the number of quanta
radiated from the part of the particle trajectory equal to the cylinder
radius. For waveguides with small radii the number of quanta radiated from
the unit length of the trajectory can be fairly large. However, in the
experimental conditions, a number of additional factors must be taken into
account. In particular, the lower limit of the distance from the cylinder
surface $r_{0}-r_{c}$, appearing in the condition (\ref{omnlim}), is
restricted by the bunch radius. It is of interest to note that for a bunch
moving in vacuum ($\varepsilon _{1}=1$) the upper limit in (\ref{omnlim})
linearly increases with increasing beam energy. For example, considering the
parameters of the bunch used in experiments of Ref. \cite{Cook09} (bunch
energy $60\,\mathrm{MeV}$ and radius $0.1\,\mathrm{mm}$) and taking $%
r_{0}-r_{c}=1\,\mathrm{mm}$, for the upper limit of the frequency $\omega
/(2\pi )$, obtained from the right-hand side of (\ref{omnlim}), we get $%
\approx 5.6\,\mathrm{THz}$.

\section{Emission of surface polaritons}

\label{sec:SP}

In this section we consider the radiation on the modes of the dielectric
cylinder with $\lambda _{j}^{2}<0$, $j=0,1$, that correspond to surface
polaritons. For the Fourier components of the fields with a given $n$, the
radial dependence is described by the function $K_{n}(|\lambda _{1}|r)$ in
the region $r>r_{c}$ and by the function $I_{n}(|\lambda _{0}|r)$ inside the
cylinder, $r<r_{c}$, and these modes correspond to surface waves. Depending
on the electromagnetic properties of the contacting media, various types of
surface waves can be excited on the separating boundary. Among them,
motivated by wide applications in light-emitting devices, surface imaging,
data storage, surface-enhanced Raman spectroscopy, biomedicine, plasmonic
solar cells, etc., the surface plasmon polaritons have attracted a great
deal of attention \cite{Agra82}. They are evanescent electromagnetic waves
propagating along a metal-dielectric interface as a result of collective
oscillations of electrons coupled to electromagnetic field. Among the most
important properties of surface plasmon polaritons is the possibility for
concentration of the fields beyond the diffraction limit that enhances the
local field strengths by several orders of magnitude. Other types of active
media instead of metals can also support surface polariton modes. Examples
are organic and inorganic dielectrics, ionic crystals, doped semiconductors
and metamaterials \cite{West10}. An important advantage of these materials
is the possibility to control the parameters in the dispersion relations for
dielectric permittivity and magnetic permeability. In particular, they can
be used for the extension of plasmonics to the infrared and terahertz
frequency ranges.

In the problem under consideration, the formula for the energy losses in the
form of surface polaritons is obtained from (\ref{Wc3}) introducing instead
of the functions $J_{n}(\lambda _{0}r_{c})$ and $J_{n\pm 1}(\lambda
_{0}r_{c})$ the modified Bessel functions $I_{n}(|\lambda _{0}|r_{c})$ and $%
I_{n\pm 1}(|\lambda _{0}|r_{c})$. Similar to the case of guided modes, we
can see that for real $k_{z}$ the integrand is real and, hence, the only
nonzero contribution to the integral comes from the poles of the integrand.
As before, the latter correspond to the zeros of the function $\alpha _{n}$.
In the case under consideration this function is written as%
\begin{equation}
\alpha _{n}=\frac{\varepsilon _{0}}{\varepsilon _{1}-\varepsilon _{0}}+\frac{%
1}{2}\sum_{l=\pm 1}\left[ 1+\frac{|\lambda _{1}|}{|\lambda _{0}|}\frac{%
I_{n+l}(|\lambda _{0}|r_{c})K_{n}(|\lambda _{1}|r_{c})}{I_{n}(|\lambda
_{0}|r_{c})K_{n+l}(|\lambda _{1}|r_{c})}\right] ^{-1}.  \label{alfn1}
\end{equation}%
The equation $\alpha _{n}=0$ determines the dispersion relation for the
surface modes (see, for example, \cite{Ashl74}). By taking into account that
the term with the ratios of the modified Bessel functions is always
positive, we conclude that the equation may have solutions if and only if $%
0<1/\left( 1-\varepsilon _{1}/\varepsilon _{0}\right) <1$, or, equivalently,
under the condition $\varepsilon _{1}/\varepsilon _{0}<0$. Hence, in order
to have eigenmodes of the cylinder with $\lambda _{0}^{2}<0$ the dielectric
permittivities of the cylinder and of the surrounding medium should have
opposite signs. Of course, this is a result that is well-known for planar
interfaces as well.

\begin{figure}[tbph]
\begin{center}
\epsfig{figure=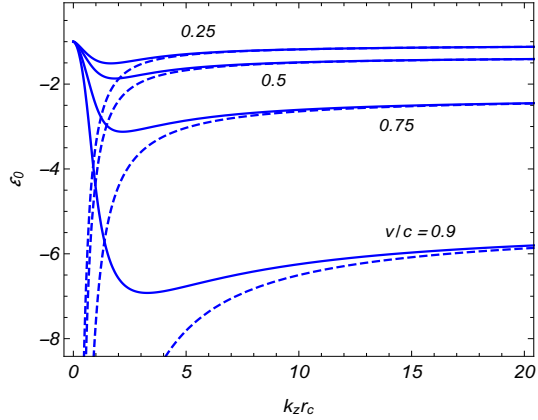,width=7.cm,height=5.5cm}
\end{center}
\caption{The localization of the eigenmodes of the cylinder with respect to $%
k_{z}r_{c}$ for $n=0$ (dashed curves) and $n=1$ (full curves). For the
surrounding medium we have taken $\protect\varepsilon _{1}=1$ and the
numbers near the curves are the values of $v/c$.}
\label{fig7}
\end{figure}

As before, we will denote by $k_{n,s}$ the eigenvalues for $k_{z}$, being
the roots of the equation $\alpha _{n}=0$. Unlike the case of guided modes,
because of monotonicity of the modified Bessel functions, the equation $%
\alpha _{n}=0$ for surface polaritons has a finite number of solutions. For
a given $n$ we can have one or two roots. This feature is illustrated in
figure \ref{fig7} where the roots with respect to $k_{z}r_{c}$ are plotted
versus $\varepsilon _{0}$ for $\varepsilon _{1}=1$ and for several values of
the ratio $v/c$ (numbers near the curves). The dashed and full curves
correspond to $n=0$ and $n=1$, respectively. By taking into account that the
product $k_{n,s}r_{c}$ depends on the parameters through the combinations $%
\varepsilon _{0}/\varepsilon _{1}$ and $\beta _{1}$, we see that figure \ref%
{fig7} describes the distribution of the roots for $\varepsilon _{1}\neq 1$
as well. In the limit $k_{z}r_{c}\rightarrow \infty $ the curves tend to the
limiting value $\varepsilon _{0}=\varepsilon _{0}^{(\infty )}$ which depends
on the ratio $v/c$ and does not depend on $n$. Below it will be shown that
\begin{equation}
\varepsilon _{0}^{(\infty )}=-\frac{\varepsilon _{1}}{1-\beta _{1}^{2}}.
\label{eps0inf}
\end{equation}
As seen from the graphs, for $n=0$ one has a single root in the region $%
\varepsilon _{0}<\varepsilon _{0}^{(\infty )}$ and there are no surface
modes in the range $\varepsilon _{0}>\varepsilon _{0}^{(\infty )}$. For $%
n\geq 1$ the surface modes are present in the region $\varepsilon _{0}^{%
\mathrm{(m)}}\leq \varepsilon _{0}<-\varepsilon _{1}$ (see the asymptotic
analysis below), where the minimal value $\varepsilon _{0}^{\mathrm{(m)}}$
depends on $n$ and $v/c$. For $\varepsilon _{0}$ close to the minimal value
one has two roots, whereas in the remaining range a single root exists. In
the limit $v/c\rightarrow 0$ one has $\varepsilon _{0}^{\mathrm{(m)}%
}\rightarrow -\varepsilon _{1}$ and for $v/c\ll 1$ the surface modes are
present in the narrow range for the permittivity $\varepsilon _{0}$ with the
length of the order $\beta _{1}^{2}$.

The distribution of the roots presented in figure \ref{fig7} can be
understood qualitatively considering the asymptotic behavior of the function
$\alpha _{n}$ from (\ref{alfn1}). For $k_{z}r_{c}\gg n+1$, assuming also
that $|\lambda _{j}|r_{c}\gg n+1$, we get%
\begin{equation}
\alpha _{n}\approx \frac{\varepsilon _{0}}{\varepsilon _{1}-\varepsilon _{0}}%
+\left( 1+\sqrt{\frac{1-\beta _{1}^{2}}{1-\beta _{0}^{2}}}\right)
^{-1}\left( 1+\frac{1}{2|\lambda _{0}|r_{c}}\right) .  \label{alfLarge}
\end{equation}%
From here it follows that for the graphs in figure \ref{fig7} one has $%
\varepsilon _{0}\rightarrow \varepsilon _{0}^{(\infty )}\equiv -\varepsilon
_{1}/(1-\beta _{1}^{2})$ in the limit $k_{z}r_{c}\rightarrow \infty $. Note
that this asymptotic does not depend on $n$. In the opposite limit of small $%
k_{z}r_{c}\ll 1$, we get%
\begin{eqnarray}
\alpha _{0} &\approx &\frac{\varepsilon _{1}}{\varepsilon _{1}-\varepsilon
_{0}}+\frac{1}{4}\left( 1-\beta _{1}^{2}\right) (k_{z}r_{c})^{2}\ln
(k_{z}r_{c}),  \notag \\
\alpha _{1} &\approx &\frac{1}{2}\frac{\varepsilon _{1}+\varepsilon _{0}}{%
\varepsilon _{1}-\varepsilon _{0}}-\frac{1}{4}\left( 1-\beta _{0}^{2}\right)
(k_{z}r_{c})^{2}\ln (k_{z}r_{c}),  \label{alfSmall1}
\end{eqnarray}%
and
\begin{equation}
\alpha _{n}\approx \frac{1}{2}\frac{\varepsilon _{1}+\varepsilon _{0}}{%
\varepsilon _{1}-\varepsilon _{0}}+k_{z}^{2}r_{c}^{2}\frac{2+\left[ \left(
n-1\right) \varepsilon _{1}-\left( n+1\right) \varepsilon _{0}\right]
v^{2}/c^{2}}{8n\left( n^{2}-1\right) },  \label{alfSmall2}
\end{equation}%
for $n>1$. From these asymptotic expressions it follows that for the roots
of the equation $\alpha _{0}=0$ we have $\varepsilon _{0}\rightarrow -\infty
$ in the limit $k_{z}r_{c}\rightarrow 0$. This feature is seen in figure \ref%
{fig7} (dashed curves). For $n\geq 1$ the asymptotic expressions (\ref%
{alfSmall1}) and (\ref{alfSmall2}) imply that for the roots of $\alpha
_{n}=0 $ one has $\varepsilon _{0}\rightarrow -\varepsilon _{1}$ in the
limit $k_{z}r_{c}\rightarrow 0$. Again, this is confirmed by figure \ref%
{fig7} (full curves).

In considerations of surface polaritons the allowance for the dispersion of
the dielectric permittivity of the cylinder, $\varepsilon _{0}=\varepsilon
_{0}\left( \omega \right) $, is required. Among the most popular models used
in surface plasmonics (see, for example, \cite{Agra82,West10}) is the Drude
type dispersion%
\begin{equation}
\varepsilon _{0}(\omega )=\varepsilon _{\infty }-\frac{\omega _{p}^{2}}{%
\omega ^{2}+i\gamma \omega },  \label{eps0}
\end{equation}%
where $\varepsilon _{\infty }$ is the background dielectric constant, $%
\omega _{p}$ is the plasma frequency and $\gamma $ is the characteristic
collision frequency or the damping coefficient. The plasma frequency can be
tuned changing the carrier concentrations in the material. For example, in
the terahertz range doped semiconductors are used. Alternatively, one can
control the electromagnetic properties by using artificially constructed
materials.

In the discussion below we will ignore the imaginary part in (\ref{eps0})
assuming that the absorption is small. In the corresponding model the
surface polaritons are radiated in the spectral range $\omega <\omega _{p}/%
\sqrt{\varepsilon _{\infty }}$. Let us consider the properties of those
modes in the asymptotic regions of the dimensionless parameter $\omega
_{p}r_{c}/v$. For $\omega _{p}r_{c}/v\ll 1$ one has $|\lambda _{j}|r_{c}\ll
1 $ and for $\alpha _{0}$ we have the asymptotic expression (\ref{alfSmall1}%
). As it has been already mentioned, from that asymptotic it follows that $%
-\varepsilon _{0}\gg 1$ or $\omega /\omega _{p}\ll 1$ for the $n=0$ modes
(for composite materials with high negative permittivity see, for example,
\cite{Yao16} and references therein). For the dispersion (\ref{eps0}) with $%
\gamma =0$, from the asymptotic expression of $\alpha _{0}$ for the
frequencies of $n=0$ surface polaritons in the range $\omega _{p}r_{c}/v\ll
1 $ one gets%
\begin{equation}
\frac{\omega }{\omega _{p}}\approx \frac{\left( \omega _{p}r_{c}/v\right)
^{-1}}{\sqrt{1-\beta _{1}^{2}}}\exp \left[ -\frac{2\varepsilon _{1}(\omega
_{p}r_{c}/v)^{-2}}{1-\beta _{1}^{2}}\right] .  \label{Smallomp}
\end{equation}%
For the modes with $n\geq 1$ and under the condition $\omega _{p}r_{c}/v\ll
1 $ we use the asymptotics (\ref{alfSmall1}) and (\ref{alfSmall2}). From
those expressions, in combination with (\ref{eps0}), it follows that one
should have $|\varepsilon _{0}+\varepsilon _{1}|\ll 1$. By taking into
account (\ref{eps0}), for the surface polariton modes with $n\geq 1$ we
obtain%
\begin{equation}
\frac{\omega }{\omega _{p}}\rightarrow \frac{1}{\sqrt{\varepsilon _{\infty
}+\varepsilon _{1}}},\;\omega _{p}r_{c}/v\rightarrow 0.  \label{Smallomp1}
\end{equation}%
In the opposite limit $\omega _{p}r_{c}/v\gg n+1$ one has $|\lambda
_{j}|r_{c}\gg n+1$ and we can use the asymptotic expression (\ref{alfLarge}%
). From the equation $\alpha _{n}=0$, in combination with $\varepsilon
_{0}/\varepsilon _{1}<0$, it follows that $\varepsilon _{0}/\varepsilon
_{1}\approx -1/(1-\beta _{1}^{2})$. For the dispersion (\ref{eps0}) this
gives
\begin{equation}
\frac{\omega ^{2}}{\omega _{p}^{2}}\approx \frac{1}{\varepsilon _{\infty
}+\varepsilon _{1}/\left( 1-\beta _{1}^{2}\right) },  \label{Largeomp}
\end{equation}%
in the asymptotic region $\omega _{p}r_{c}/v\gg n+1$.

\begin{figure}[tbph]
\begin{center}
\begin{tabular}{cc}
\epsfig{figure=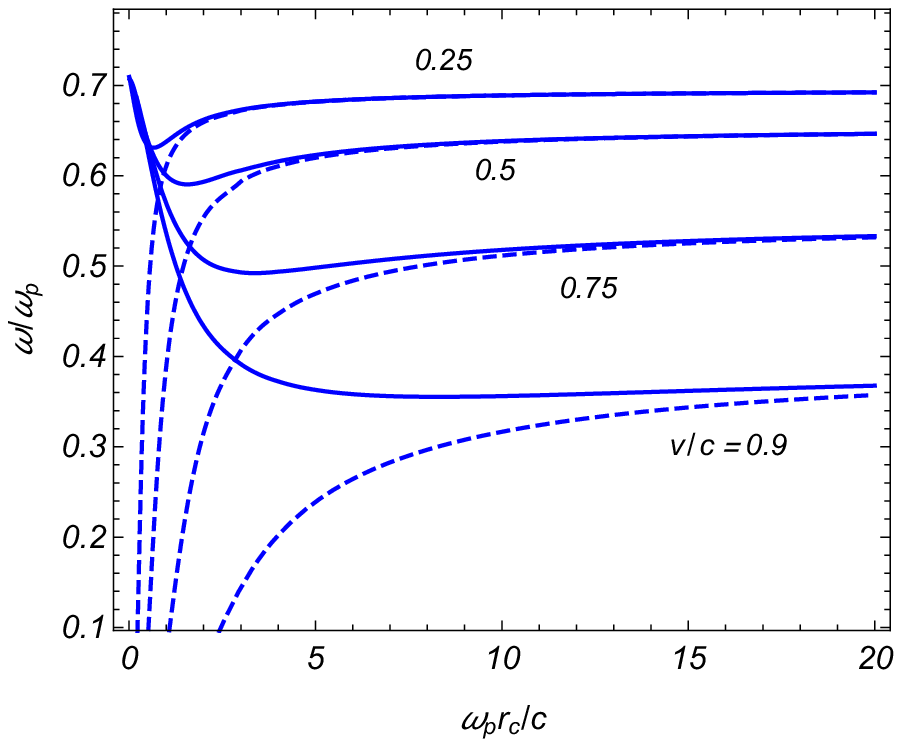,width=7.cm,height=5.5cm} & \quad %
\epsfig{figure=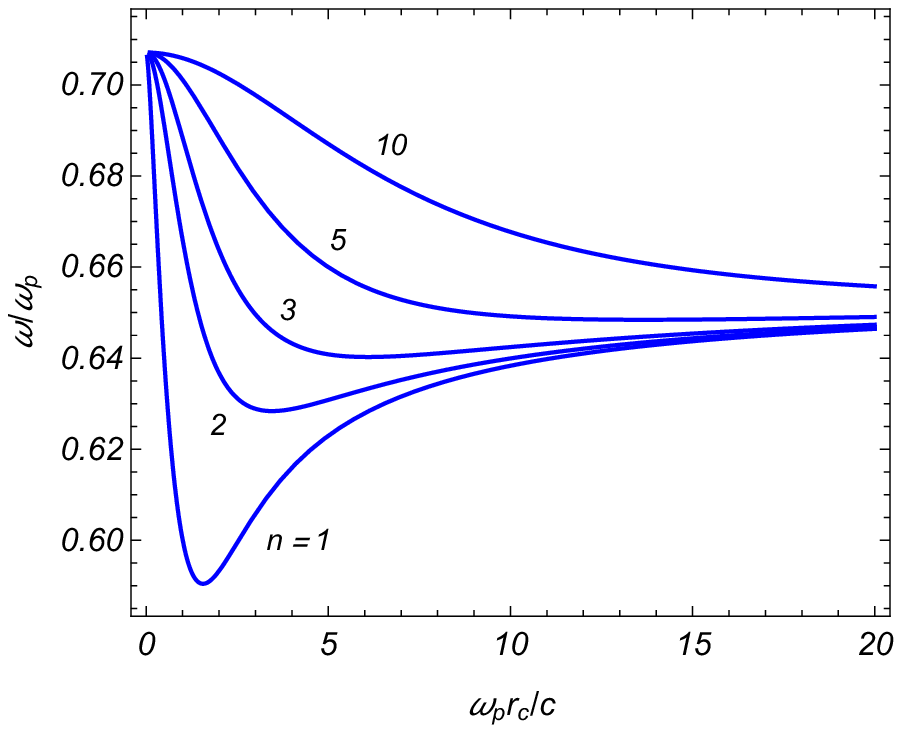,width=7.cm,height=5.5cm}%
\end{tabular}%
\end{center}
\caption{Eigenfrequencies of the cylinder corresponding to surface
polaritons versus $\protect\omega _{p}r_{c}/c$ for the dispersion law (%
\protect\ref{eps0}) with $\protect\gamma =0$ and for $\protect\varepsilon %
_{1}=1$. On the left panel the full and dashed curves correspond to the
modes with $n=1$ and $n=0$, respectively, and the numbers near the curves
are the values of the ratio $v/c$. The graphs on the right panel are plotted
for $v/c=0.5$ and the numbers near the curves correspond to the values of $n$%
.}
\label{fig8}
\end{figure}

In the left panel of figure \ref{fig8}, for dispersion law (\ref{eps0}) with
$\varepsilon _{\infty }=1$ and $\gamma =0$, we present the frequencies for
the eigenmodes of the cylinder as functions of the plasma frequency. The
full and dashed curves correspond to $n=1$ and $n=0$ respectively. The
numbers near the curves are the values of ratio $v/c$. The right panel of
figure \ref{fig8} presents the frequencies of the eigenmodes for different
values of $n$ (numbers near the curves). From the data plotted in figure \ref%
{fig8} we see that for $n=0$ the frequencies of surface polaritons are in
the range
\begin{equation}
\omega <\omega _{p}\left( \varepsilon _{\infty }+\frac{\varepsilon _{1}}{%
1-\beta _{1}^{2}}\right) ^{-1/2}.  \label{omlim}
\end{equation}%
For the frequencies of the modes with $n\geq 1$, in addition to the upper
limit in (\ref{omlim}) one has a lower limit: $\omega \geq \omega ^{(m)}$.
The limiting frequency increases with increasing $n$ and tends to $\omega
_{p}/\sqrt{\varepsilon _{\infty }+\varepsilon _{1}/(1-\beta _{1}^{2})}$ for
large values of $n$.

Having clarified the distribution of the eigenmodes we turn to the radiation
intensity for surface polaritons. Similar to the case of guided modes, in
order to specify the integration contour near the poles of the integrand in (%
\ref{Wc3}), we introduce an imaginary part of the permittivity $\varepsilon
_{0}$ and use the expansion (\ref{alfExp}). The poles are located at $%
k_{z}=k_{n,s}-i\varepsilon _{0}^{\prime \prime }b_{n,s}$, where $b_{n,s}$ is
defined by (\ref{bns}). We have checked numerically that for $\lambda
_{0}^{2}<0$ one has $\partial _{\varepsilon _{0}^{\prime }}\alpha
_{n}(k_{n,s},\varepsilon _{0}^{\prime })>0$. From here it follows that the
poles $k_{z}=k_{n,s}$ should be avoided from above for $\sigma _{n,s}>0$ and
from below for $\sigma _{n,s}<0$ by small semicircles in the complex plane $%
k_{z}$. The energy radiated per unit time is presented as (\ref{Ins}), where
the radiation intensity for surface polaritons of the angular frequency $%
\omega _{n,s}=vk_{n,s}$ is expressed as%
\begin{equation}
I_{n,s}=2\delta _{n}q^{2}\frac{v}{\varepsilon _{1}}\sqrt{1-\beta _{1}^{2}}%
k_{z}^{2}\frac{K_{n}^{2}(|\lambda _{1}|r_{0})}{V_{n}^{K}}\frac{%
I_{n}(|\lambda _{0}|r_{c})}{r_{c}|\alpha _{n}^{\prime }(k_{z})|}\sum_{p=\pm
1}\left. \frac{I_{n+p}(|\lambda _{0}|r_{c})}{V_{n+p}^{K}}\right\vert
_{k_{z}=k_{n,s}},  \label{InsSP}
\end{equation}%
where $|\lambda _{j}|=k_{z}\sqrt{1-\beta _{j}^{2}}$,
\begin{equation}
V_{n}^{F}=I_{n}(|\lambda _{0}|r_{c})\partial _{r_{c}}F_{n}(|\lambda
_{1}|r_{c})-F_{n}(|\lambda _{1}|r_{c})\partial _{r_{c}}I_{n}(|\lambda
_{0}|r_{c}),  \label{VnF}
\end{equation}%
for $F=I,K$. Note that one has $V_{n}^{K}<0$. Similar to the case of the
guided modes, the radiation intensity is suppressed by the factor $%
e^{-2|\lambda _{1}|(r_{0}-r_{c})}$ for the modes with $|\lambda
_{1}|r_{c}\gg 1$. Unlike the guided modes, there is no velocity threshold
for the generation of surface polaritons.

Let us consider asymptotic estimates of the radiation intensity for the
dispersion relation (\ref{eps0}) with $\gamma =0$. In accordance with the
analysis given above, in the limit $v\rightarrow 0$ one has $\omega
\rightarrow \omega _{p}/\sqrt{\varepsilon _{1}+\varepsilon _{\infty }}$. By
taking into account that $|\lambda _{j}|r_{c}\approx \omega r_{c}/v$, we see
that the arguments of the modified Bessel functions in (\ref{InsSP}) are
large. By using the corresponding asymptotic expressions we conclude that in
the limit $v\rightarrow 0$ the radiation intensity is suppressed by the
factor $\exp [-2\omega _{p}\left( r_{0}-r_{c}\right) /(v\sqrt{\varepsilon
_{1}+\varepsilon _{\infty }})]$. Now we turn to the behavior of the
radiation intensity in the limiting regions of the combination $\omega
_{p}r_{c}/v$. In the region $\omega _{p}r_{c}/v\ll 1$ and for the modes $n=0$
we get $I_{0,s}\propto \left( \omega /\omega _{p}\right) ^{2}/\left( \omega
_{p}r_{c}/v\right) ^{2}$, where the ratio $\omega /\omega _{p}$ is given by (%
\ref{Smallomp}). The corresponding radiation intensity is exponentially
small. For the surface modes with $n\geq 1$ the radiation intensity in the
same region is estimated as
\begin{equation}
I_{n,s}\approx \frac{2q^{2}v}{r_{c}^{2}}\frac{\left( r_{c}/r_{0}\right)
^{2n}\left( \omega _{p}r_{c}/v\right) ^{2}}{n\left( \varepsilon _{\infty
}+\varepsilon _{1}\right) ^{2}}.  \label{InSmallomp}
\end{equation}%
The corresponding frequencies are given by (\ref{Largeomp}) and the
radiations intensity is suppressed by the factor $\left( \omega
_{p}r_{c}/v\right) ^{2}$. In the opposite limit, $\omega _{p}r_{c}/v\gg n+1$%
, the radiation intensity is estimated as
\begin{equation}
I_{n,s}\approx \frac{2q^{2}\omega _{p}}{r_{0}}\frac{\left( \omega /\omega
_{p}\right) ^{3}}{1-\beta _{1}^{2}/2}\exp \left[ -2(r_{0}-r_{c})\frac{\omega
}{v}\sqrt{1-\beta _{1}^{2}}\right] ,  \label{InLargeomp}
\end{equation}%
with the radiation frequency from (\ref{Largeomp}). Similar to the case of
the guided modes, the frequency of the radiated surface polaritons is
restricted by the condition (\ref{omnlim}). For the values of the bunch
parameters discussed at the end of the previous section the upper limit of
the frequency for surface polaritons is of the order of $10\,\mathrm{THz}$.
On the other hand, our consideration is restricted by the condition $\omega
\gg \gamma $ that is required to neglect the imaginary part of the
dielectric permittivity in (\ref{eps0}). For metals the ratio $\gamma /(2\pi
)$ is of the order of $10\,\mathrm{THz}$ and the approximation used in
deriving (\ref{InsSP}) is not valid for the abovementioned values of the
bunch characteristics. Note that the formula (\ref{W2}) for the energy
losses is valid for general case of the complex function $\varepsilon
_{0}(\omega )$. In the presence of imaginary part of $\varepsilon
_{0}(\omega )$, in addition to the radiation part, $dW/dz$ contains also
other types of the energy losses.

\begin{figure}[tbph]
\begin{center}
\epsfig{figure=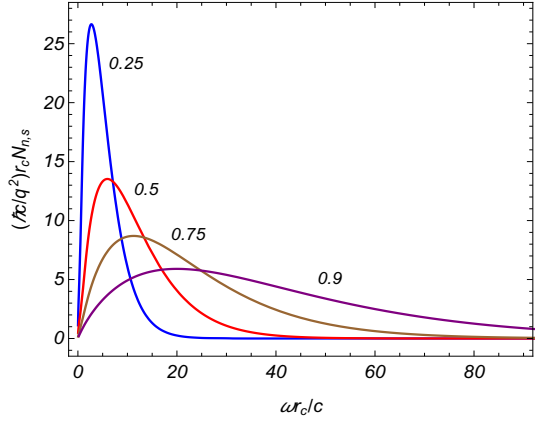,width=7.cm,height=5.5cm}
\end{center}
\caption{The spectral distribution of the number of radiated surface
polaritons on the modes with $n=0$ for a cylinder immersed in the vacuum.
The numbers near the curves are the values for $v/c$.}
\label{fig9}
\end{figure}

In figure \ref{fig9} we have displayed the number of the radiated quanta for
surface polaritons as a function of the frequency for the modes with $n=0$
and for $\varepsilon _{1}=1$, $r_{0}/r_{c}=1.05$. The numbers near the
curves are the values of the ratio $v/c$. Note that different frequencies
correspond to different values of the permittivity $\varepsilon _{0}$. The
value for $\varepsilon _{0}$ corresponding to given frequency can be found
from the data depicted in figure \ref{fig7}. We see that the number of the
radiated quanta is large enough compared to the case of the radiation of
guided modes.

Here, a comment is in order. In the numerical evaluations corresponding to
figure \ref{fig9}, for a given value of $\varepsilon _{0}$, with fixed $%
\varepsilon _{1}$ and $v$, we solve the equation $\alpha _{n}=0$ with
respect to $k_{z}r_{c}$. At this step, for a given $\varepsilon _{0}$, the
specific form of the dispersion is not required. The latter is needed in the
numerical evaluation of the radiation intensity. Indeed, the radiation
intensity contains the derivative $\alpha _{n}^{\prime }(k_{z})$. By taking
into account the relation $\omega =k_{z}v$, in the expression for $\alpha
_{n}^{\prime }(k_{z})$ the derivative $\partial _{\omega }\varepsilon
_{0}(\omega )$ will enter coming from the terms in (\ref{alfn1}) with $%
\lambda _{0}$ and from the first term in the right-hand side. Hence, for the
evaluation of the radiation intensity on a given frequency $\omega $, in
addition to $\varepsilon _{0}(\omega )$, the value of the derivative $%
\partial _{\omega }\varepsilon _{0}(\omega )$ is required. Plotting the
graphs in figure \ref{fig9} we have assumed that the dispersion is weak and
the part of the derivative $\alpha _{n}^{\prime }(k_{z})$ containing $%
\partial _{\omega }\varepsilon _{0}(\omega )$ has been ignored. In the
spectral range with $\varepsilon _{0}<0$ this idealization may lead to
problems. For example, a problem appears in the evaluation of the radiation
intensity on the mode $n=1$. In the absence of dispersion there exists a
special value of $\omega r_{c}/c$ (or equivalently of $\varepsilon _{0}$)
for which the derivative $\alpha _{n}^{\prime }(k_{z})$ becomes zero. This
means that the corresponding point is a higher order pole of the integrand
in (\ref{Wc3}). One of possible ways to regularize this singularity is to
include the imaginary part of the permittivity $\varepsilon _{0}(\omega )$.
Note that this kind of problem does not appear in the problem of radiation
from a charge circulating around a cylinder, discussed in \cite{Kota19}. The
reason is that in the latter problem, for a given $n$, the radiation
frequency $\nu =n/T$, with $T$ being the charge rotation period, and $k_{z}$
are independent variables. As a consequence of this, for evaluation of $%
\alpha _{n}^{\prime }(k_{z})$ the derivative $\partial _{\omega }\varepsilon
_{0}(\omega )$ is not required and a given value of $\varepsilon _{0}$
determine both the eigenvalues of $k_{z}$ and the radiation intensity.

\begin{figure}[tbph]
\begin{center}
\begin{tabular}{cc}
\epsfig{figure=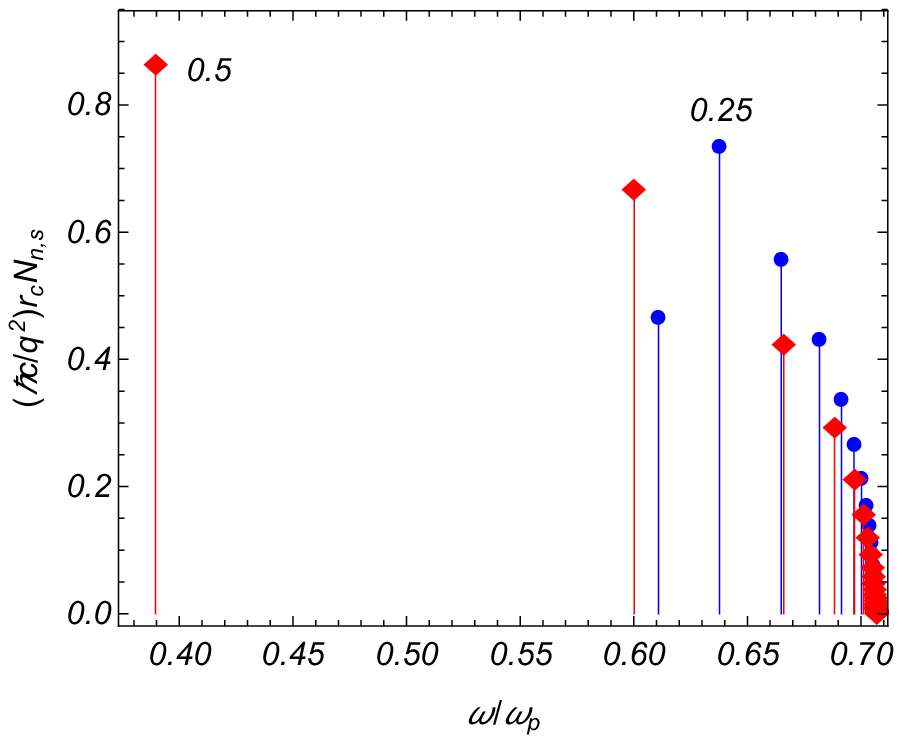,width=7.cm,height=5.5cm} & \quad %
\epsfig{figure=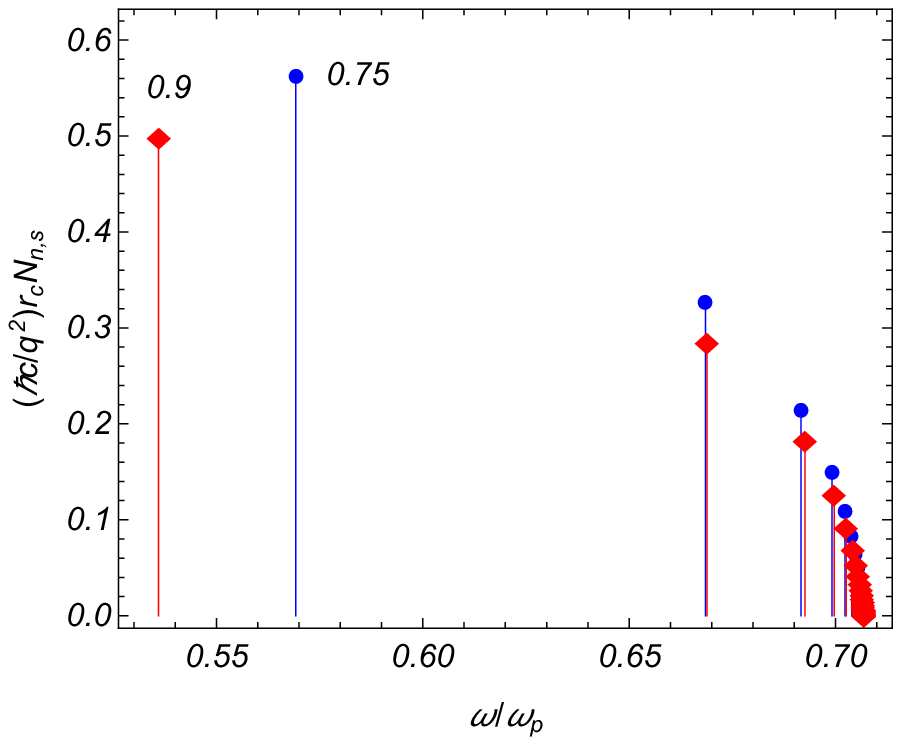,width=7.cm,height=5.5cm}%
\end{tabular}%
\end{center}
\caption{The number of the radiated quanta in the form of surface
polaritons, as a function of the frequency, for different values of $n\in
[0,20]$ (for the values of the parameters see the text).}
\label{fig10}
\end{figure}

\begin{figure}[tbph]
\begin{center}
\begin{tabular}{cc}
\epsfig{figure=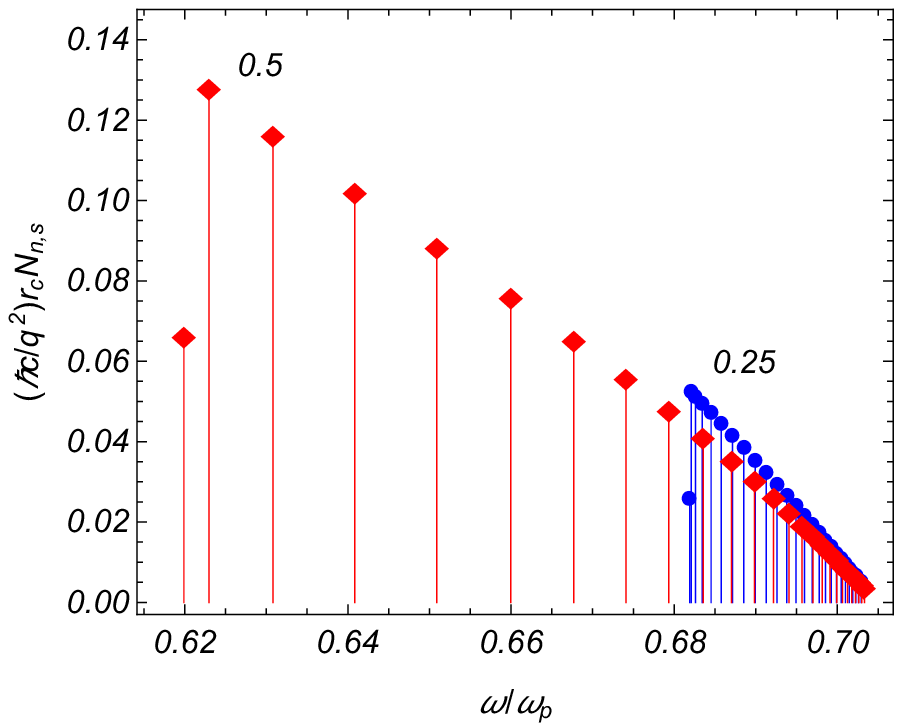,width=7.cm,height=5.5cm} & \quad %
\epsfig{figure=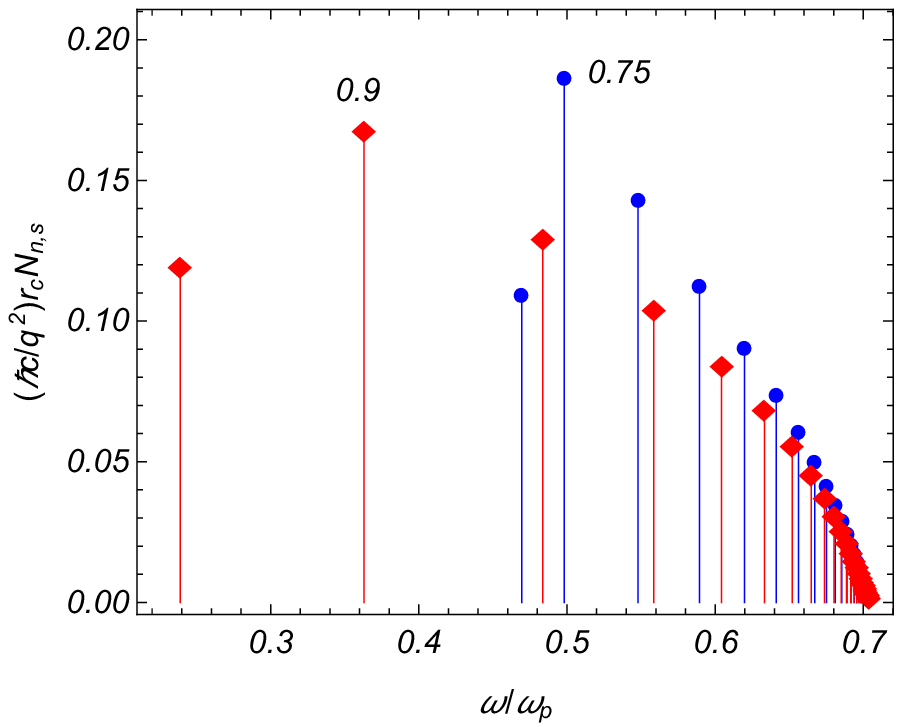,width=7.cm,height=5.5cm}%
\end{tabular}%
\end{center}
\caption{The same as in figure \protect\ref{fig10} for $\protect\omega %
_{p}r_{c}/c=5$.}
\label{fig11}
\end{figure}

Given the importance of dispersion in discussing the emission of surface
polaritons, in figure \ref{fig10}, for the dispersion law (\ref{eps0}) with $%
\varepsilon _{\infty }=1$, $\gamma =0$, and for $\varepsilon _{1}=1$, the
number of the radiated quanta for surface polaritons is presented as a
function of the frequency (in units of the plasma frequency) $\omega /\omega
_{p}=\omega _{n,s}/\omega _{p}$ for $n\in \lbrack 0,20]$. In the numerical
evaluation we have taken $r_{0}/r_{c}=1.05$ and $\omega _{p}r_{c}/c=1$. The
plot markers circles and diamonds correspond to $v/c=0.25$ and $v/c=0.5$ on
the left panel and to $v/c=0.75$ and $v/c=0.9$ on the right panel. Note that
the eigenfrequencies increase with increasing $n$, $\omega _{n,s}<\omega
_{n+1,s}$. The same data for $\omega _{p}r_{c}/c=5$ are presented in figure %
\ref{fig11}. As it already has been concluded from the asymptotic analysis,
for large $n$ the radiation frequencies tend to the value $\omega _{p}/\sqrt{%
\varepsilon _{\infty }+\varepsilon _{1}}$ ($=\omega _{p}/\sqrt{2}$ for the
examples in figures \ref{fig10} and \ref{fig11}). The spectral range of the
radiated surface polaritons becomes narrower with decreasing $v/c$.

In the discussion above we have considered the radiation from a single point
charge. The corresponding results for the spectral density of the radiation
intensity can be generalized for a bunch containing $N_{q}$ particles. Let
us consider a simple case of the bunch with transverse beam size smaller
than the radiation wavelength. The $z$-component of the current density is
presented as $j_{3}^{\mathrm{(b)}}(t,\mathbf{r})=\sum_{m=1}^{N_{q}}j_{m3}(t,%
\mathbf{r})$, where the expression for the current density $j_{m3}(t,\mathbf{%
r})$ for the $m$th particle in the bunch is obtained from (\ref{jz}) by the
replacement $z\rightarrow z-z_{m}$, with $z_{m}$ being the $z$-coordinate of
the $m$th particle at the initial moment $t=0$. The expressions for the
Fourier components of the fields are obtained from the corresponding
formulas given above for a single charge by adding the factor $%
\sum_{m=1}^{N_{q}}e^{-ik_{z}z_{m}}$ with $k_{z}=\omega /v$. In the
expression for the radiation intensity the factor $\left\vert
\sum_{m=1}^{N_{q}}e^{-ik_{z}z_{m}}\right\vert ^{2}$ will appear. The double
sum in this modulus squared, $\sum_{m,m^{\prime }=1}^{N_{q}}$, is decomposed
into the incoherent contribution with $m^{\prime }=m$ and the remaining
coherent contribution. Introducing the longitudinal distribution function of
the bunch $f(z)$ in accordance with $\sum_{m=1}^{N_{q}}\int_{-\infty
}^{+\infty }dz\,\delta (z-z_{n})e^{-ik_{z}z}=N_{q}\int_{-\infty }^{+\infty
}dz\,f(z)e^{-ik_{z}z}$, we see that the radiation intensity from a bunch is
obtained from the formulas for a single charge by adding an additional
geometrical factor
\begin{equation}
N_{q}\left[ 1+(N_{q}-1)\left\vert g(k_{z})\right\vert ^{2}\right] ,
\label{BunchFac}
\end{equation}%
where $g(k_{z})=\int_{-\infty }^{+\infty }dz\,f(z)e^{-ik_{z}z}$. For a
Guassian bunch one has $f(z)=e^{-z^{2}/2\sigma _{z}^{2}}/(\sqrt{2\pi }\sigma
_{z})$ and $\left\vert g(k_{z})\right\vert ^{2}=\exp (-k_{z}^{2}\sigma
_{z}^{2})$, with $\sigma _{z}$ being the rms bunch length. The second term
in the squared brackets of (\ref{BunchFac}) presents the contribution of the
coherent effects in the radiation intensity.

In the discussion above we have assumed that the waveguide has infinite
length along its axis. This allowed to provide an exact solution for the
problem under consideration. In fact, the most part of the papers cited
above, that consider the radiation in waveguides, uses this approximation.
The results for the fields given above will approximate the features for a
dielectric cylinder with the length $L_{c}$ under the conditions $%
r,r_{0},\lambda _{\mathrm{r}}\ll L_{c}$, with $\lambda _{\mathrm{r}}$ being
the radiation wavelength. For a finite cylinder, in addition to the
radiations discussed above there will be diffraction radiation at the ends.
Another interesting effect at the termination of the waveguide corresponds
to the transformation of guided modes and surface polaritons to free
electromagnetic fields propagating in surrounding medium.

\section{Conclusion}

\label{sec:Conc}

We have investigated the radiation emitted by a charge uniformly moving
outside a dielectric cylinder, parallel to its axis. The electric and
magnetic fields are found for general cases of dielectric permittivities of
the cylinder and surrounding medium. First we have investigated the spectral
density for the CR intensity in the exterior medium by evaluating the energy
flux at large distances from the charge. The spectral density is given by (%
\ref{Ifl3}) with functions $f_{n}^{(p)}$ from (\ref{fnz}). It has been shown
that the influence of the cylinder on the CR is essentially different in the
cases $\varepsilon _{0}<\varepsilon _{1}$ and $\varepsilon _{0}>\varepsilon
_{1}$. The characteristic feature in the first case is presented in figure %
\ref{fig2} with relatively small oscillations of the spectral density of the
radiation intensity around the value corresponding to the radiation in a
homogeneous medium. For wavelengths much smaller than the cylinder diameter
the oscillations enter into quasiperiodic regime. These oscillations result
from the interference of the direct CR, the CR reflected from the cylinder
and also the CR formed inside the cylinder if the corresponding Cherenkov
condition is obeyed. In the case $\varepsilon _{0}>\varepsilon _{1}$ strong
narrow peaks may appear in the spectral distribution of the radiation
intensity. We have specified the conditions for the presence of those peaks.
They come from the terms of the series in (\ref{Ifl3}) with large values of $%
n$ and are closely related to the eigenvalue equation for the dielectric
cylinder. The equation (\ref{Peaks}) that determines the spectral locations
of the peaks is obtained form the eigenvalue equation $\alpha _{n}=0$
ignoring the exponentially small terms of the order $|J_{n}(\lambda
_{1}r_{c})|/Y_{n}(\lambda _{1}r_{c})$. Under the Cherenkov condition with
the dielectric permittivity of the surrounding medium the eigenvalue
equation has no solutions and the radiation modes corresponding to the
strong peaks could be called as "quasimodes" of the dielectric cylinder. The
radiation on this types of the modes may also appear in the spectral range
where $\varepsilon _{0}<-\varepsilon _{1}$. We have analytically estimated
the heights and widths of the peaks by using the asymptotic expressions for
the cylinder functions for large arguments.

If the Cherenkov condition for the exterior medium is not satisfied,
depending on the spectral range, two types of radiations may appear
propagating inside the cylindrical waveguide. They have discrete spectrum
determined by the dispersion relation $\alpha _{n}=0$. The corresponding
fields exponentially decay as functions of the distance from the cylinder
surface and they correspond to guided modes and to surface polaritons. For
guided modes $\lambda _{0}^{2}>0$ and the Cherenkov condition is satisfied
for the dielectric permittivity of the cylinder. For those modes the radial
dependence of the fields is expressed in terms of the Bessel function $%
J_{n}(\lambda _{0}r)$ and the radiation intensity is given by (\ref{Wc3b}).
The lower threshold for the guided modes frequency increases with increasing
$n$ and the radiation frequency range is determined by (\ref{omnlim}).

Unlike the guided modes, there is no velocity threshold for the emmision of
surface polaritons. They are radiated in the spectral range where the
dielectric permittivities of the cylinder and of the surrounding medium have
opposite signs. The corresponding radial dependence of the radiation fields
inside the cylinder is described by the Bessel modified function $%
I_{n}(|\lambda _{0}|r)$ and the radiation intensity on a given frequency is
expressed as (\ref{InsSP}). The dispersion for surface polaritons is
qualitatively different for the modes with $n=0$ and $n\geq 1$. For $n=0$
there is a upper threshold for the values of the permittivity $\varepsilon
_{0}$ (given by (\ref{eps0inf})): the eigenvalue equation has a single root
in the region $\varepsilon _{0}<\varepsilon _{0}^{(\infty )}$ and there are
no surface modes in the range $\varepsilon _{0}>\varepsilon _{0}^{(\infty )}$%
. In the case $n\geq 1$, a single or two surface modes exist in the finite
range $\varepsilon _{0}^{\mathrm{(m)}}\leq \varepsilon _{0}<-\varepsilon
_{1} $ with the lower threshold $\varepsilon _{0}^{\mathrm{(m)}}$ depending
on $n$ and $v/c$. In the nonrelativistic limit $\varepsilon _{0}^{\mathrm{(m)%
}}$ tends to $-\varepsilon _{1}$ and the surface modes are present in the
narrow range for $\varepsilon _{0}$ with the length of the order $\beta
_{1}^{2}$. For illustration of general results, as an example of dispersion
for dielectric permittivity of the cylinder we have considered Drude type
model. In the limiting regions of the dimensionless parameter $\omega
_{p}r_{c}/v$ the frequencies of the surface modes are estimated by (\ref%
{Smallomp})-(\ref{Largeomp}). The radiation intensities for surface
polaritons in those regions are approximated by (\ref{InSmallomp}) and (\ref%
{InLargeomp}). The spectral range of the generated surface polaritons
becomes narrower with decreasing $v/c$. Having the fields and radiation
intensity for a single charge, one can obtain the corresponding result for a
bunch of particles. In the simple case of the bunch with small transverse
size, the effect of the bunch appears in the form of the geometrical factor (%
\ref{BunchFac}) determined by the bunch longitudinal form-factor.

\section*{Acknowledgement}

This work has been supported by Grant No. 18T-1C397 of the Committee of
Science of the Ministry of Education, Science, Culture and Sport RA.

\end{document}